\documentclass[%
reprint,
superscriptaddress,
altaffilletter,
nofootinbib,
 amsmath,amssymb,
 aps,
 prd,
]{revtex4-1}

\usepackage{times} %
\usepackage{mathtools}
\usepackage{graphicx}%
\usepackage{dcolumn}%
\newcolumntype{d}[1]{D{.}{.}{#1}}
\usepackage{bm}%
\usepackage{verbatim} %
\usepackage{cprotect} %
\usepackage{url}
\usepackage{subfigure} %
\usepackage{tikz}
\usepackage{afterpage}
\usepackage{nicefrac}
\usepackage{tabularx}
\newcolumntype{L}{>{\raggedright\arraybackslash}X}
\usepackage{multirow}
\usepackage[english]{babel}
\usepackage{natbib} %
\usepackage{glossaries} %
\usepackage{hyperref}%
\usepackage{aas_macros} %

\clearpage{}%
\newacronym{GW}{GW}{gravitational-wave}
\newacronym{LIGO}{LIGO}{Laser Interferometer Gravitational-wave Observatory}
\newacronym{ECO}{ECO}{exotic compact object}
\newacronym{IMR}{IMR}{inspiral-merger-ringdown}
\newacronym{IMRE}{IMRE}{inspiral-merger-ringdown-echo}
\newacronym{MLE}{MLE}{maximum likelihood estimator}
\newacronym{MAP}{MAP}{maximum a posteriori estimator}
\newacronym{O1}{O1}{the first observing run}
\newacronym{ROC}{ROC}{receiver operating characteristic}\clearpage{}%

\newcommand{\this}{paper} %
\newcommand{\DocumentID}{P1800319}
\newcommand{\etal}{\textit{et al.}}

\clearpage{}%
\newcommand{\GaussianEfficiencyPercentage}{82.3}
\newcommand{\GaussianEfficiencyDecimal}{0.823}

\newcommand{\OOneEfficiencyPercentage}{61.1}
\newcommand{\OOneEfficiencyDecimal}{0.611}

\newcommand{\GaussianOneSigma}{-0.9}
\newcommand{\GaussianTwoSigma}{-0.4}
\newcommand{\GaussianThreeSigma}{1.1}
\newcommand{\GaussianFourSigma}{1.5}
\newcommand{\GaussianFiveSigma}{1.9}

\newcommand{\OOneOneSigma}{0.1}
\newcommand{\OOneTwoSigma}{1.5}
\newcommand{\OOneThreeSigma}{4.0}
\newcommand{\OOneFourSigma}{5.4}
\newcommand{\OOneFiveSigma}{5.7}

\newcommand{\GaussianThreshold}{\GaussianFiveSigma}
\newcommand{\OOneThreshold}{\OOneFiveSigma}\clearpage{}%

\graphicspath{{figures/}}

\usepackage{xcolor} %
\begin{document}

\title{Template-based gravitational-wave echoes search using Bayesian model selection}%

\author{Rico K. L. Lo}
\email{kllo@caltech.edu}
\affiliation{Department of Physics, The Chinese University of Hong Kong, Shatin, New Territories, Hong Kong
}
\affiliation{LIGO Laboratory, California Institute of Technology, Pasadena, CA 91125, US
}

\author{Tjonnie G. F. Li}
\affiliation{Department of Physics, The Chinese University of Hong Kong, Shatin, New Territories, Hong Kong
}

\author{Alan J. Weinstein}
\affiliation{LIGO Laboratory, California Institute of Technology, Pasadena, California 91125, USA
}

\date{April 30, 2019}%

\begin{abstract}
The ringdown of the gravitational-wave signal from a merger of two black holes has been suggested as a probe of the structure of the remnant compact object, which may be more exotic than a black hole. It has been pointed out that there will be a train of echoes in the late-time ringdown stage for different types of exotic compact objects. In this \this{}, we present a template-based search methodology using Bayesian statistics to search for echoes of gravitational waves. Evidence for the presence or absence of echoes in gravitational-wave events can be established by performing Bayesian model selection. The Occam factor in Bayesian model selection will automatically penalize the more complicated model that echoes are present in gravitational-wave strain data because of its higher degree of freedom to fit the data. We find that the search methodology was able to identify gravitational-wave echoes with Abedi \etal{}'s echoes waveform model about \GaussianEfficiencyPercentage{}\% of the time in simulated Gaussian noise in the Advanced LIGO and Virgo network and about \OOneEfficiencyPercentage{}\% of the time in real noise in the first observing run of Advanced LIGO with $\geq 5\sigma$ significance.
Analyses using this method are performed on the data of Advanced LIGO's first observing run, and we find no statistical significant evidence for the detection of gravitational-wave echoes. In particular, we find $< 1\sigma$ combined evidence of the three events in Advanced LIGO's first observing run.
The analysis technique developed in this \this{} is independent of the waveform model used, and can be used with different parametrized echoes waveform models to provide more realistic evidence of the existence of echoes from exotic compact objects.

\end{abstract}

\maketitle

\section{Introduction}
As of this writing, the \gls{LIGO} \cite{2015CQGra..32g4001L} and Advanced Virgo \cite{2015CQGra..32b4001A} have successfully detected ten compact binary coalescence events from binary black hole systems \cite{PhysRevLett.116.061102, PhysRevLett.116.241103, PhysRevLett.118.221101, 2017arXiv171105578T, 2017PhRvL.119n1101A, GWTC1Catalog} and one binary neutron star collision \cite{2017PhRvL.119p1101A}. These discoveries mark the beginning of a new era of \gls{GW} astronomy and astrophysics, where we can infer and probe the properties and structure of astronomical objects using gravitational waves.

During the inspiral stage of gravitational-wave emission from the coalescence of a compact binary system, for instance a binary black hole system, the two black holes spiral towards each other with an increasing orbital frequency. Eventually, they coalesce in the merger stage to form one single black hole. The final black hole then relaxes to a Kerr black hole  during the ringdown stage.

Cardoso, Franzin and Pani \cite{2016PhRvL.116q1101C} first pointed out that the ringdown part of the gravitational-wave signal can be used as a probe of the structure of a compact object. A very compact object, not necessarily a black hole, with a light ring will also exhibit a similar ringdown as that of a black hole. Cardoso, Hopper, Macedo, Palenzuela and Pani \cite{2016PhRvD.94h4031C} further showed that a similar ringdown stage will also be exhibited for different types of \emph{\glspl{ECO}} with a light ring (or a photon sphere), and there will be a train of echoes in the late-time ringdown stage associated with the photon sphere. Examples of \glspl{ECO} are theoretical alternatives to black holes, such as gravastars and fuzzballs. A common feature of these alternatives is that there is some kind of structure near the would-be event horizon. The echoes in the late-time ringdown stage are caused by repeated and damped reflections between the effective potential barrier and the reflective structure.
Cardoso \etal{} also showed that the time delay between each echo $\Delta t_{\text{echo}}$ can be used to infer the nature of an \gls{ECO} \cite{2016PhRvD.94h4031C}, namely
\begin{equation}\label{Eq: Theoretical delta t_echo}
\Delta t_{\text{echo}} \sim - nM \log \left( \frac{l}{M} \right),
\end{equation}
where $M$ is the mass of the \gls{ECO}, $l \ll M$ is the microscopic correction of the location of the \gls{ECO} surface from the Schwarzschild radius, and $n$ is an integer of the order of 1 which depends on the nature of the \gls{ECO}.

Abedi, Dykaar and Afshordi published a paper in December 2016, claiming that they had found tentative evidence of Planck-scale structure near the black hole event horizons at a combined $2.9\sigma$ significance level \cite{2017PhRvD..96h2004A} of GW150914, LVT151012 and GW151226 using the matched filtering technique. However, their analysis methodology, especially the estimation of statistical significance, was questioned \cite{2016arXiv161205625A, 2017arXiv170103485A}. Various teams have also proposed methods to estimate the parameters of the gravitational-wave echoes \cite{2017PhRvD..96f4045M, 2018PhRvD..97l4037W}, and to search for echoes in a morphology-agnostic way\cite{2017arXiv171206517C, 2018PhRvD..98b4023T}.

In this \this{}, we present a template-based search methodology using Bayesian inference to search for echoes of gravitational waves in compact binary coalescence events.
The analysis technique in this \this{} can be used with different gravitational-wave echoes waveform models to provide robust evidence of the existence of echoes from \glspl{ECO} by showing consistent results using different models. Detecting an exotic compact object would be a groundbreaking discovery as this would revolutionize our understanding of compact objects, and that this can only be achieved by gravitational-wave observations. In parallel to this work, there are efforts to search for gravitational-wave echoes using Bayesian model selection with templates using the inference package \texttt{PyCBC Inference}\cite{2018arXiv181104904N}.

This \this{} is structured as follows. In Sec. \ref{Sec: Methods}, we first establish the methodology of the search, namely Bayesian model selection and parameter estimation in Sec. \ref{Subsec: Bayesian hypothesis testing}, the gravitational-wave echoes template model in Sec. \ref{Subsec: templates} and statistical significance estimation in \ref{Subsec: statistical significance}. We then describe ways to evaluate the sensitivity of a search in Sec. \ref{Subsec: Evaluation of search sensitivity}, and the combination of Bayesian evidence from multiple gravitational-wave echoes events in Sec. \ref{Subsec: Combine evidence}. In Sec. \ref{Sec: Results}, we first describe our implementation in Sec. \ref{Subsec: Implementation}, and then we present the results of a Bayesian parameter estimation and model selection of the presence of echoes versus their absence that were performed on simulated data with Gaussian noise in Sec. \ref{Subsec: PE on simulated data} and Sec. \ref{Model selection on simulated data} respectively. Then we evaluate the performance of the search in simulated Gaussian noise and real noise in \gls{O1} in Sec. \ref{Subsec: search sensitivity, efficiency and accuracy of simulated data}. We demonstrate the idea of combining multiple gravitational-wave echoes events in Sec. \ref{Subsec: Demonstration of combining evidence}. Finally in Sec. \ref{Subsec: O1 Search Results}, we show the search results for the three events in \gls{O1}.

\section{Methods}\label{Sec: Methods}
\subsection{Bayesian model selection and parameter estimation}\label{Subsec: Bayesian hypothesis testing}
To search for echoes of gravitational waves from the coalescence of exotic compact objects, we perform Bayesian model selection analyses on confirmed gravitational-wave events. Here we consider two hypotheses $\mathcal{H}_{0}$ and $\mathcal{H}_{1}$, which can also be considered as the null hypothesis and alternative hypothesis in the frequentist language, and they are
\begin{align*}
\mathcal{H}_{0} \coloneqq & \, \text{No echoes in the data} \Rightarrow d = n + h_{\text{IMR}},\\
\mathcal{H}_{1} \coloneqq & \, \text{There are echoes in the data} \Rightarrow d = n + h_{\text{IMRE}},\\
\end{align*}
where $d$ denotes the \acrlong{GW} data, $n$ denotes the instrumental noise and $h_{\text{IMR}}$, $h_{\text{IMRE}}$ denote the \gls{IMR} gravitational-wave signal and \gls{IMRE} gravitational-wave signal respectively. Note that we \textit{assume there is a gravitational-wave signal in the data} since we perform the search after the gravitational-wave signal has been identified, and we are only interested in knowing whether there are echoes in the data or not. When the null hypothesis $\mathcal{H}_{0}$ is true, that means the data contain a \gls{GW} signal with \gls{IMR}. When the alternative hypothesis $\mathcal{H}_{1}$ is true instead, that means the data contain a \gls{GW} signal with both echoes and an \acrlong{IMR} part.

In the context of gravitational-wave data analysis, suppose that the strain data $d(t)$ from a detector only consist of noise $n(t)$, which we assume to be Gaussian and stationary (we will relax these assumptions in what follows). The probability that the noise $n(t)$ has a realization $n_{0}(t)$ (with zero mean) is given by \cite{MaggioreMichele2007GWV1}
\begin{equation}\label{Eq: Gaussian noise}
p(n_{0}) = \mathcal{N} \exp \left[ -\dfrac{1}{2} \int_{-\infty}^{+\infty} df \dfrac{|\tilde{n_{0}}(f)|^{2}}{(1/2)S_{n}(f)} \right],
\end{equation}
where $\mathcal{N}$ is a normalization constant and $S_{n}(f)$ is the power spectrum density of noise.
We introduce the notion of \emph{noise-weighted inner product}, namely
\begin{equation}\label{Eq: Noise-weighted inner product}
\langle A | B \rangle = 4\Re \int_{f_{\text{low}}}^{f_{\text{high}}} df \dfrac{\tilde{A}^{*}(f) \tilde{B}(f)}{S_{n}(f)},
\end{equation}
where $f_{\text{low}}$ and $f_{\text{high}}$ are the low-frequency cutoff and high-frequency cutoff respectively. The integration is performed over a finite range because detectors are taking samples at a finite rate, and hence there is a theoretical upper limit on the maximum frequency that one can resolve from the data, and detectors are not sensitive enough below some frequency threshold.
Using the inner product, we can rewrite Eq. \ref{Eq: Gaussian noise} as
\begin{equation}
p(n_{0}) = \mathcal{N} \exp \left[ -\dfrac{1}{2} \langle n_{0}|n_{0} \rangle \right].
\end{equation}

Now, suppose the strain data $d(t)$ consist of both noise $n_{0}(t)$ and a \gls{GW} signal modeled by a template $h(t; \vec{\theta})$, where $\vec{\theta}$ is a set of parameters of the template that describe the signal, that is
\begin{equation}
n_{0}(t) = d(t) - h(t; \vec{\theta}).
\end{equation}
Then the likelihood $p(d|\vec{\theta}, \mathcal{H}, I)$ for a single detector can be obtained from Eq. \ref{Eq: Gaussian noise}:
\begin{equation}\label{Eq: Single detector likelihood}
p(d|\vec{\theta}, \mathcal{H}, I) = \mathcal{N} \exp \left[ -\dfrac{1}{2} \langle d(t) - h(t; \vec{\theta})|d(t) - h(t; \vec{\theta}) \rangle \right],
\end{equation}
where $I$ denotes the knowledge known prior to the selection; in this case we knew prior to the model selection that the data contain a \gls{GW} signal.
For the case of multiple detectors (for example, H1, L1 and V1), if we assume that the noise distributions for each detector are all Gaussian and stationary, and more importantly independent of each other, then we have
\begin{widetext}
\begin{equation}\label{Eq: Multiple detectors likelihood}
p(d_{\text{H1}}, d_{\text{L1}}, d_{\text{V1}}|\vec{\theta}, \mathcal{H}, I) = \prod_{i \in \{\text{H1}, \text{L1}, \text{V1} \}} \mathcal{N}_{i} \exp \left[ -\dfrac{1}{2} \langle d_{i}(t_{i}) - h(t_{i}; \vec{\theta})|d_{i}(t_{i}) - h(t_{i}; \vec{\theta}) \rangle \right].
\end{equation}
\end{widetext}

With the notion of noise-weighted inner product, we can also define the matched-filtering \emph{signal-to-noise ratio} (SNR) $\rho$, which tells us how strong a signal is with respect to the noise, as follows:
\begin{equation}\label{Eq: SNR}
\rho^2 = \frac{\langle d|h \rangle^{2}}{\langle h|h \rangle},
\end{equation}
where $d$ denotes the gravitational-wave data and $h$ is a gravitational-wave signal template. If we have multiple detectors (for example, H1, L1 and V1), we can define the \emph{network SNR} squared as the sum of the matched filtering SNR squared in each detector
\begin{equation}\label{Eq: Network SNR}
\rho^2_{\text{network}} = \sum_{i \in \{\text{H1}, \text{L1}, \text{V1} \}} \rho^{2}_{i}.
\end{equation}
In the optimal case of a template that exactly matches the signal in the data, the matched-filtering SNR is bound by the \emph{optimal SNR}, which is given by
\begin{equation}\label{Eq: Optimal SNR}
\rho^2_{\text{optimal}} = \frac{\langle h|h \rangle^{2}}{\langle h|h \rangle},
\end{equation}
and can be used as an indication of how strong a signal is.

In Bayesian model selection, we compute the Bayes factor $\mathcal{B}_{{0}}^{{1}}$ and odds ratio $\mathcal{O}_{{0}}^{{1}}$, which are defined as
\begin{align}\label{Eqs: Bayes factors and odds ratio}
\mathcal{B}_{{0}}^{{1}} & = \frac{p(d|\mathcal{H}_{1}, I)}{p(d|\mathcal{H}_{0}, I)}, \\
\mathcal{O}_{{0}}^{{1}} & = \frac{p(\mathcal{H}_{1}|d, I)}{p(\mathcal{H}_{0}|d, I)} = \mathcal{B}_{{0}}^{{1}} \times \frac{p(\mathcal{H}_{1}|I)}{p(\mathcal{H}_{0}|I)}.
\end{align}
In the Bayesian language, the odds ratio has the interpretation that when $\mathcal{O}_{{0}}^{{1}} > 1$, it means that the data favor the hypothesis $\mathcal{H}_{1}$, and vice versa. For the sake of simplicity, we will drop the superscript and subscript on the Bayes factor $\mathcal{B}$ and odds ratio $\mathcal{O}$ from now on when the context is clear. If we assume that each hypothesis is equally likely prior to the model selection, namely
\begin{equation}
p(\mathcal{H}_{0} \,| I) = p(\mathcal{H}_{1} \,| I) = \frac{1}{2},
\end{equation}
then the odds ratio is simply the Bayes factor, that is
\begin{equation}\label{Eq: Echo Bayes factor}
\mathcal{O} = \mathcal{B} = \frac{p(d|\mathcal{H}_{1}, I)}{p(d|\mathcal{H}_{0}, I)}.
\end{equation}

It is often more convenient to work in log space, namely we compute the log posterior, log likelihood and log prior. We take the natural logarithm on both sides of Eq. \ref{Eq: Echo Bayes factor}, and we have
\begin{align}\label{Eq: Echo log Bayes factor}
& \ln \mathcal{O} = \ln \mathcal{B} \nonumber \\
 &= \ln p(d|\mathcal{H}_{1}, I) - \ln p(d|\mathcal{H}_{0}, I) \nonumber \\
& = \ln {Z}_{1} - \ln {Z}_{0},
\end{align}
where the term $Z_{i} \equiv p(d|\mathcal{H}_{i}, I)$, which is known as the \emph{evidence} for the hypothesis $\mathcal{H}_{i}$, can be estimated by numerically integrating over the template parameter space $\vec\theta_i$ of hypothesis $\mathcal{H}_{i}$ using a sampling algorithm such as parallel-tempering Markov chain Monte Carlo with thermodynamic integration \cite{10.2307/2280232, 10.2307/2334940, PhysRevLett.57.2607, doi:10.1063/1.1751356} or nested sampling \cite{doi:10.1063/1.1835238}. In this paper, we will use nested sampling (or more specifically \texttt{LALInferenceNest} \cite{2015PhRvD..91d2003V}). Apart from estimating the evidence, we can also obtain a set of posterior samples as byproducts of the nested sampling algorithm, which allow us to perform parameter estimation \footnote{A detailed discussion of parameter estimation can be found in Ref. \cite{2015PhRvD..91d2003V}.} with little additional computational cost. We can calculate various estimators of parameters as point estimates from the posterior samples, such as the \gls{MLE}, which is 
\begin{equation}
\hat{\vec{\theta}}_{i, \text{MLE}} = \arg\max \mathcal{L}(\vec{\theta}_{i} |d, \mathcal{H}_{i}, I),
\end{equation}
where $\mathcal{L}(\vec{\theta}_{i} |d, \mathcal{H}_{i}, I) = p(d|\vec{\theta}_{i}, \mathcal{H}_{i}, I)$ is the likelihood as a function of the parameters $\vec{\theta}_{i}$. Another estimator is the \gls{MAP}, which is
\begin{equation}
\hat{\vec{\theta}}_{i, \text{MAP}} = \arg\max p(\vec{\theta_{i}}|d, \mathcal{H}_{i}, I),
\end{equation}
where $p(\vec{\theta}_{i} |d, \mathcal{H}_{i}, I)$ is the posterior distribution of parameters $\vec{\theta}_{i}$.

To obtain the evidence for the hypothesis $\mathcal{H}_{i}$ in the context of gravitational-wave data analysis, we use gravitational-wave waveform templates that assume $\mathcal{H}_{i}$ being true to compute the log likelihood.

\subsubsection{Occam factor}\label{Subsubsec: Occam's factor}
One must be cautious when performing model selection that the model which fits the data best does not imply that the model gives the highest evidence. A more complicated model, i.e. with more free parameters, is more easily affected by noise in the data than a simpler model, i.e. with less free parameters. This is similar to overfitting in regression. Suppose there are $N$ data points for fitting; one can always use a degree $N - 1$ polynomial to fit all points, but very likely the fitted polynomial will not generalize well to new data because it was affected by the noise in the data.

Bayesian analysis embodies the \emph{Occam factor} and penalizes more complicated models automatically. To illustrate this idea, suppose there are two hypotheses, namely $\mathcal{H}_{0}$ and $\mathcal{H}_{1}$. Without loss of generality, we assume that $\dim \vec{\theta}_{1} > \dim \vec{\theta}_{0}$, where $\dim \vec{\theta}_{i}$ denotes the dimension of the parameter vector $\vec{\theta}_{i}$ that describes the hypothesis $\mathcal{H}_{i}$. If the posterior distribution has a sharp peak at $\vec{\theta}_{i} = \vec{\theta}_{i, \text{MAP}}$ with width $\sigma_{i, \text{posterior}}$, then the integral for evidence $Z_{i}$ can be approximated using Laplace's method. We first write the integral for evaluating the evidence of the hypothesis $\mathcal{H}_{i}$ into the standard form for Laplace's method
\begin{equation}\label{Eq: Evidence integral}
Z_{i} = \int \exp \left\{ \ln \left[ p(d|\vec{\theta}_{i}, \mathcal{H}_{i}, I) p(\vec{\theta}_{i}|\mathcal{H}_{i}, I) \right] \right\} d\vec{\theta}_{i}.
\end{equation}
Let $f(\vec{\theta}_{i}) \equiv \ln \left[ p(d|\vec{\theta}_{i}, \mathcal{H}_{i}, I) p(\vec{\theta}_{i}|\mathcal{H}_{i}, I) \right]$, and we expand $f(\vec{\theta}_{i})$ about the sharp peak $\vec{\theta}_{i} = \vec{\theta}_{i, \text{MAP}}$, which gives
\begin{equation}
f(\vec{\theta}_{i}) = f(\vec{\theta}_{i, \text{MAP}}) + \frac{f^{''}(\vec{\theta}_{i, \text{MAP}})}{2} (\vec{\theta}_{i} - \vec{\theta}_{i, \text{MAP}})^2 + \mathcal{O}(\vec{\theta}_{i} - \vec{\theta}_{i, \text{MAP}})^3,
\end{equation}
where the first derivative $f^{'}$ vanishes and the second derivative $f^{''}(\vec{\theta}_{i, \text{MAP}}) < 0$ at the local maximum.
Substituting this back into Eq. \ref{Eq: Evidence integral}, we have
\begin{equation}\label{Eq: Approx evidence integral}
Z_{i} \approx \exp f(\vec{\theta}_{i, \text{MAP}}) \int \exp \left[ - \frac{\lvert f^{''}(\vec{\theta}_{i, \text{MAP}}) \rvert}{2} (\vec{\theta}_{i} - \vec{\theta}_{i, \text{MAP}})^2 \right] d\vec{\theta}_{i},
\end{equation}
where the integral becomes a Gaussian integral in the limit that the integration is performed over $(-\infty,\infty)$.

Finally we can approximate the evidence $Z_{i}$ (up to some constant factors) by
\begin{equation}\label{Eq: Laplace approximation of evidence}
Z_{i} \approx p(d|\vec{\theta}_{i, \text{MAP}}, \mathcal{H}_{i}, I) p(\vec{\theta}_{i, \text{MAP}}|\mathcal{H}_{i}, I) \sigma_{i, \text{posterior}}.
\end{equation}
Note that we have assumed that the posterior width $\sigma_{i, \text{posterior}}$ is much smaller than the width of the integration limits such that the integral in Eq. \ref{Eq: Approx evidence integral} can be well approximated by a Gaussian integral. It should also be noted that Eqs. \ref{Eq: Approx evidence integral} and \ref{Eq: Laplace approximation of evidence} \emph{were not used in our analyses}, and they were derived for the purpose of illustrating the Occam factor only.

For a more complicated model, more parameters are needed to describe the observed data. For example, for our hypothesis $\mathcal{H}_1$ (i.e., there are echoes in the data) we need to introduce extra parameters (discussed in the next section) such as the time delay between each echo $\Delta t_{\text{echo}}$ in the model selection analysis. Suppose the prior distribution of parameters $\vec{\theta}_{i}$ for each hypothesis is uniform over a width $\sigma_{i, \text{prior}}$ such that
\begin{equation}
p(\vec{\theta}_{i}|\mathcal{H}_{i}, I) = \begin{cases}
\dfrac{1}{\sigma_{i, \text{prior}}} & \text{within the range,}\\
0 & \text{otherwise.}
\end{cases}
\end{equation}
The ratio $\nicefrac{\sigma_{i, \text{posterior}}}{\sigma_{i, \text{prior}}}$ hence serves as a penalty to down-weigh the evidence $Z_{1}$ of the more complicated model $\mathcal{H}_{1}$ which has a larger prior volume, i.e. $\sigma_{1, \text{prior}} > \sigma_{0, \text{prior}}$ to account for the uncertainty of the extra parameters. This ratio, sometimes referred to as \emph{Occam factor} \cite{MacKay:2002:ITI:971143}, allows the analysis to bias the less complicated \gls{IMR}-only model in a natural way.

\subsection{Phenomenological waveform model of echoes}\label{Subsec: templates}
In this \this{}, we use the phenomenological waveform model of echoes proposed by Abedi \etal{} in Ref. \cite{2017PhRvD..96h2004A} to search for echoes of gravitational waves. It should be noted that the methodology we propose here is \emph{independent of the gravitational-wave echoes templates we used}, and different parametrized waveform models can be readily used instead of the model by Abedi \etal{} when more physical models become available in the future \cite{PhysRevD.96.084002, 2017PTEP.2017g1E01N, 2018PhRvD..97h4030C, 2018PhRvD..97l4044W, 2018PhRvD..98d4018T}. Their model was motivated by the numerical results in Ref. \cite{2016PhRvD.94h4031C}. There are five free parameters in their waveform model, with the phase change between each echo due to the reflection on an ECO surface being $\pi$. The descriptions of these five parameters are tabulated in Table \ref{Table:Abedi}.

\begin{table}[ht]
\begin{center}
\begin{ruledtabular}
\begin{tabularx}{\linewidth}{cL}
Parameter & Description \\ \hline
$\Delta t_{\text{echo}}$ & The time interval between each echo \\
$t_{\text{echo}}$ & The time of arrival of the first echo \\
$t_{0}$ & The time of truncation of the GW IMR template \\ & $\mathcal{M}_{\text{I}}(t)$ to produce the echo template $\mathcal{M}_{\text{TE,I}}(t)$ \\
$\gamma$ &  The damping factor \\
$A$ & The amplitude of the first echo relative to the IMR part\\ & of the template\\
\end{tabularx}
\end{ruledtabular}
\caption{\label{Table:Abedi}The five free parameters and the corresponding descriptions of the phenomenological gravitational-wave echoes waveform model proposed by Abedi et al.~\cite{2017PhRvD..96h2004A}. In particular, $\Delta t_{\text{echo}}$ is of the most astrophysical interest because it encapsulates the compactness of the exotic compact object that we are observing as shown in Eq. \ref{Eq: Theoretical delta t_echo}. Physically $\Delta t_{\text{echo}}$ is related to the distance between the effective potential barrier and the reflective surface that gravitational-wave echoes need to travel. Also, $A$ can tell us the typical strength of the gravitational-wave echoes emitted from exotic compact objects.}
\end{center}
\end{table}

\begin{figure}[!ht]
\begin{center}
\includegraphics[width=1.0\columnwidth]{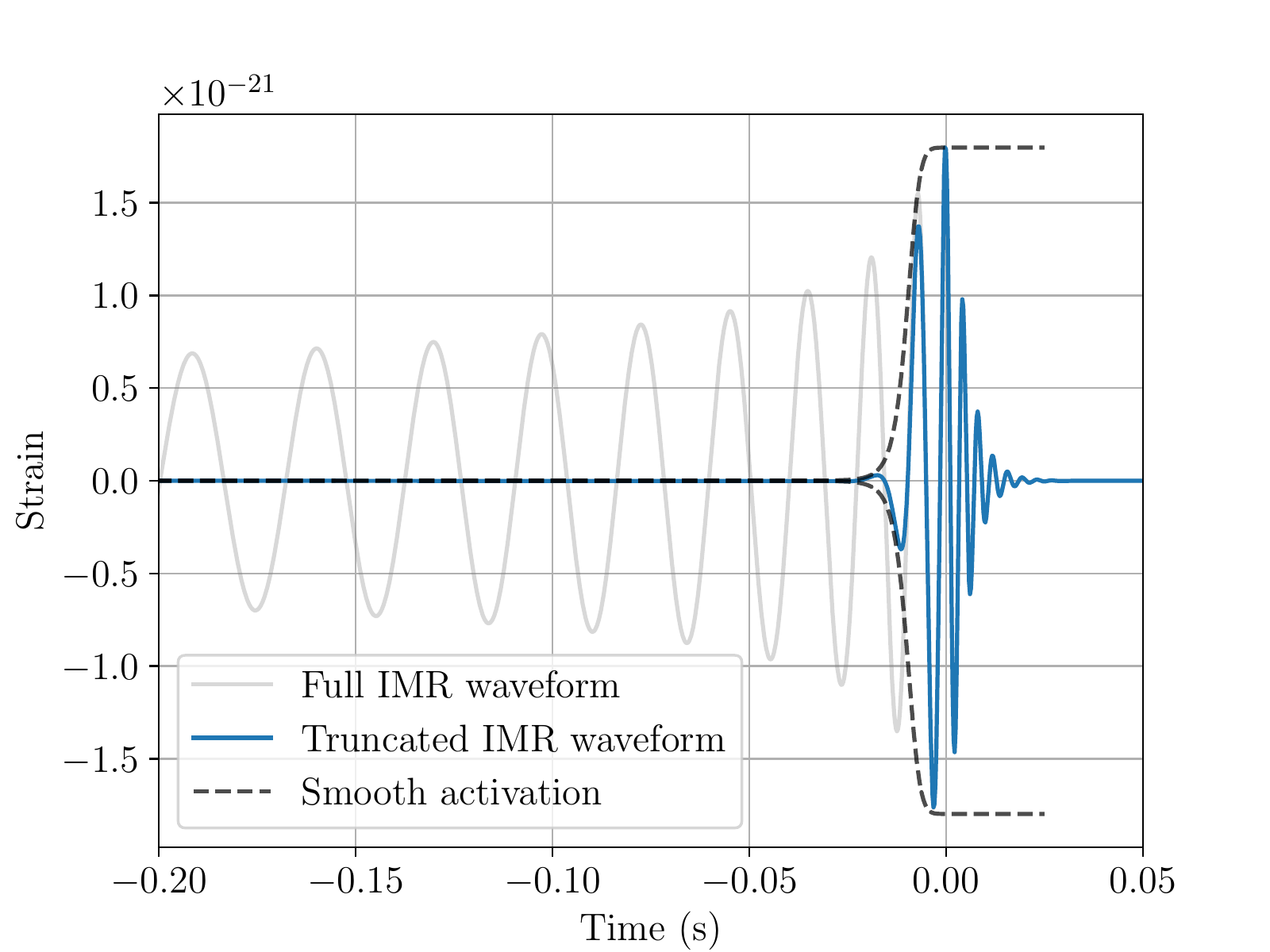}\label{Fig: Truncated h(t)}
\includegraphics[width=1.0\columnwidth]{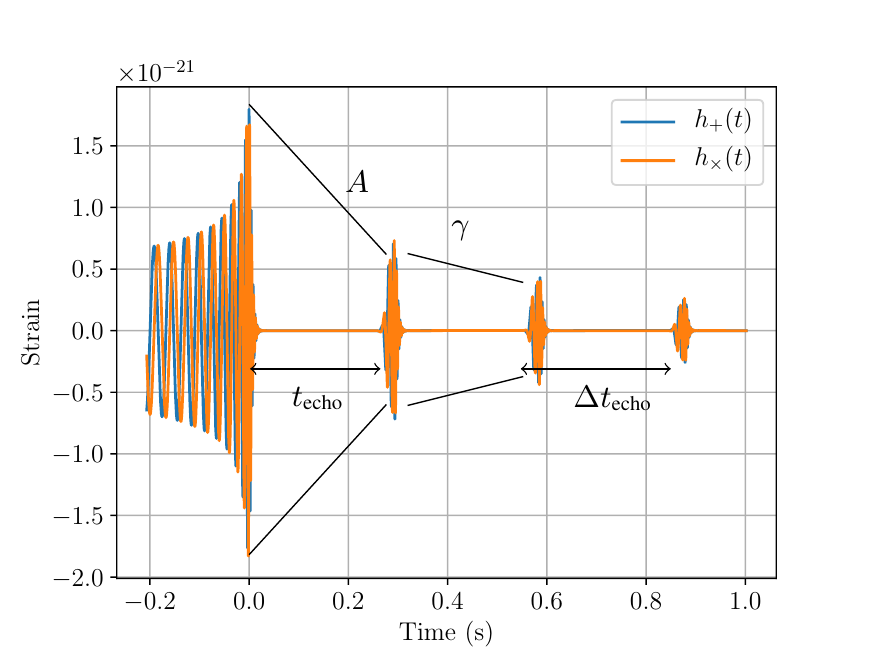}
\caption{\label{Fig: GW150914 with echo Abedi}\textit{Top panel}: To generate a template of a gravitational-wave echo (in blue), we truncate the ringdown part of the \gls{IMR} part (in gray) of a waveform by applying the smooth activation function $\Theta(t, t_{0})$ (in black dashed lines) to get the truncated IMR waveform. \textit{Bottom panel}: A plot of an \gls{IMRE} template generated using the phenomenological waveform model of echoes proposed by Abedi \etal{} We see that the first echo, which is the truncated IMR template (shown in the top panel) scaled by the parameter $A$, starts at $t_{\text{echo}}$ after the merger. Subsequent echoes, which are further scaled down by the parameter $\gamma$ due to the energy loss when the echo reflects off an ECO surface, are separated from each other in time by $\Delta t_{\text{echo}}$.}
\end{center}
\end{figure}

Using the notations in Ref. \cite{2017PhRvD..96h2004A}, the echo template $\mathcal{M}_{\text{TE,I}}(t)$ in the time domain is given by
\begin{align}\label{Eq:Abedi template}
\mathcal{M}_{\text{TE,I}}(t) & \equiv A \sum_{n = 0}^{\infty}(-1)^{n+1}\gamma^{n} \nonumber \\
& \times \mathcal{M}_{\text{T,I}}(t + t_{\text{merger}} - t_{\text{echo}} - n\Delta t_{\text{echo}}, t_{0}),
\end{align}
where $t_{\text{merger}}$ is the time of merger \footnote{Or equivalently time of coalescence, denoted by $t_{c}$.} and $\mathcal{M}_{\text{T,I}}(t)$ is a smooth activation of the GW \gls{IMR} template given by
\begin{align}
\mathcal{M}_{\text{T,I}}(t) & \equiv \Theta(t, t_{0})\mathcal{M}_{\text{I}}(t) \nonumber \\
& \equiv \frac{1}{2} \left\lbrace 1 + \tanh \left[ \frac{1}{2}\omega_{\text{I}}(t)(t - t_{\text{merger}} - t_{0}) \right] \right\rbrace \mathcal{M}_{\text{I}}(t),
\end{align}
where $\omega_{\text{I}}(t)$ denotes the angular frequency evolution of the IMR waveform as a function of time, and $\mathcal{M}_{\text{I}}(t)$ is the IMR waveform. The smooth activation $\Theta(t, t_{0})$ essentially selects the ringdown, which is the part of a waveform that one might expect to see in echoes \cite{2017PhRvD..96h2004A}. Note that the time of merger $t_{\text{merger}}$ is the only time reference, and therefore we measure all time-related echo parameters $t_{0}$, $t_{\text{echo}}$ and $\Delta t_{\text{echo}}$ \textit{with respect to $t_{\text{merger}}$}. The top and bottom panels of Fig. \ref{Fig: GW150914 with echo Abedi} show a truncated IMR time-domain waveform $\mathcal{M}_{\text{T,I}}(t)$ used to generate the echo template and a GW150914-like IMRE time-domain waveform with three echoes, respectively.

In particular, the parameter that represents the time interval between each successive echo $\Delta t_{\text{echo}}$ is of the most astrophysical interest because it encapsulates the compactness of the exotic compact object that we are observing as shown in Eq. \ref{Eq: Theoretical delta t_echo}. Physically $\Delta t_{\text{echo}}$ is related to the distance between the effective potential barrier and the reflective surface that gravitational-wave echoes need to travel. The relative amplitude $A$ can also tell us the typical strength of the gravitational-wave echoes emitted from exotic compact objects.

\subsection{Detection statistic and background estimation for statistical significance}\label{Subsec: statistical significance}
In this \this{}, the log Bayes factor $\ln \mathcal{B}$ [Eqs. \ref{Eqs: Bayes factors and odds ratio} and \ref{Eq: Echo log Bayes factor}] is the \emph{detection statistic} to decide whether we claim there is an IMRE signal or an IMR signal in data. If the log Bayes factor $\ln \mathcal{B}_{{0}}^{{1}}$, or equivalently log odds ratio $\ln \mathcal{O}_{{0}}^{{1}}$, is greater than 0, we can conclude, from the Bayesian point of view, that the data favor the alternative hypothesis that the data contain an IMRE signal more than the null hypothesis that the data contain an IMR signal, thus serving the function of distinguishing which hypothesis is more supported by the data.

After we have obtained a detection statistic, a natural question to ask is how statistically significant the detection statistic is. Simply put, how likely is it that the detection is actually caused by an IMRE signal but not due to noise? In the Bayesian school, there are different empirical scales, such as Jeffreys' scale, to interpret the strength of the Bayes factor. However, they are subjective and not universally applicable. Therefore, we are not going to use any of them in this \this{}.

Calculating the posterior probability of a hypothesis is certainly better than using a subjective scale to determine the strength of the Bayes factor. However, the Bayesian posterior probability fails to tell us the probability that the evidence is simply due to random background noise, since we only consider one set of data. The frequentist approach can answer the following question: given the null hypothesis $\mathcal{H}_{0}$ is true, what is the probability that the data are going to be \textit{as extreme as or more extreme than} the observed data? The probability that we are looking for is exactly the frequentist $p$-value. We can also interpret this $p$-value as the \emph{false-alarm probability}.

The $p$-value, which we denote as simply $p$, is related to the \emph{null distribution} of detection statistic $\ln \mathcal{B}$ by
\begin{align}\label{Eq: p-value}
p & = \Pr(\ln \mathcal{B} \geq \ln \mathcal{B}_{\text{detected}}|\mathcal{H}_{0}) \\
\nonumber & = 1 - \int_{-\infty}^{\ln \mathcal{B}_{\text{detected}}} p(\ln \mathcal{B}|\mathcal{H}_{0}) d\ln \mathcal{B},
\end{align}
where $\ln \mathcal{B}_{\text{detected}}$ is the detection statistic obtained in an analysis on a segment of data, and $p(\ln \mathcal{B}|\mathcal{H}_{0})$ is called the null distribution of $\ln \mathcal{B}$, i.e., the distribution of $\ln \mathcal{B}$ given that $\mathcal{H}_{0}$ is true.

Hence, from the null distribution, we can compute the detection statistic threshold $\ln \mathcal{B}_{\text{threshold}}$ corresponding to a certain statistical significance, e.g. $5\sigma$ and hence we can claim a detection of gravitational-wave echoes if the detection statistic of a candidate exceeds or is equal to the predetermined threshold.

\subsection{Evaluation of search sensitivity}\label{Subsec: Evaluation of search sensitivity}
Apart from getting the statistical significance of a particular candidate event of gravitational-wave echoes, we are also interested in investigating the sensitivity and accuracy of this search methodology using Bayesian model selection.

\subsubsection{Sensitive parameter space}\label{Subsubsec: Sensitive parameter space}
To quantify the sensitivity of a search, one can compute the fraction of the parameter space of echo parameters that the search can determine whether the data contain echoes or not, given a threshold on the detection statistic $\ln \mathcal{B}_{\text{threshold}}$ and gravitational-wave detectors operating at specific sensitivities. If a search is sensitive, then it should be able to cover a reasonable fraction of parameter space possible for astrophysical exotic compact objects, which is schematically defined as
\begin{equation}
\{ \vec{\theta}_{\text{echoes}} = (A, \gamma, t_{0}, t_{\text{echo}}, \Delta t_{\text{echo}}) \mid \ln \mathcal{B}(\vec{\theta}_{\text{echo}}) \geq  \ln \mathcal{B}_{\text{threshold}} \},
\end{equation}
so that the search is able to detect the existence of echoes in the data with echo parameters in the sensitive parameter space.

\subsubsection{Search efficiency}\label{Subsubsec: Search efficiency}
Another method to quantify the search sensitivity is to compute the probability that the existence of echoes will be detected given a detection statistic threshold, which is also known as the \emph{efficiency} $\zeta$. It is defined as
\begin{equation}\label{Eq: Efficiency}
\zeta = \int_{\ln \mathcal{B}_{\text{threshold}}}^{\infty} p(\ln \mathcal{B}|\mathcal{H}_{1}) d\ln \mathcal{B},
\end{equation}
where $p(\ln \mathcal{B}|\mathcal{H}_{1})$ is the \emph{foreground distribution}, i.e. the distribution of $\ln \mathcal{B}$ given that $\mathcal{H}_{1}$ is true. If a search is sensitive, then it should have a high value of efficiency $\zeta$.

\subsection{Combining Bayesian evidence from a catalog of detection events}\label{Subsec: Combine evidence}
Bayesian model selection provides us a natural way to combine evidence of the existence of exotic compact objects from multiple detection events of gravitational-wave echoes. In the following analysis, we do not assume GW events are described by the same set of echo parameters. Suppose now we have a catalog of $N_{\text{cat}}$ independent events so that we have a set of $N_{\text{cat}}$ data denoted by $d = \{ d_1, d_2, ..., d_{N_{\text{cat}}} \}$; the odds ratio for the catalog of sources is given by
\begin{align}
\mathcal{O}_{{0}}^{{1}} & = \frac{p(d_1, d_2, ..., d_{N_{\text{cat}}}|\mathcal{H}_{1}, I)p(\mathcal{H}_{1}|I)}{p(d_1, d_2, ..., d_{N_{\text{cat}}}|\mathcal{H}_{0}, I)p(\mathcal{H}_{0}|I)} \nonumber \\
& = {^{\text{(cat)}}{\mathcal{B}_{0}^{1}}} \times \frac{p(\mathcal{H}_{1}|I)}{p(\mathcal{H}_{0}|I)},
\end{align}
where 
\begin{equation}
{^{\text{(cat)}}{\mathcal{B}_{0}^{1}}} = \frac{p(d_1, d_2, ..., d_{N_{\text{cat}}}|\mathcal{H}_{1}, I)}{p(d_1, d_2, ..., d_{N_{\text{cat}}}|\mathcal{H}_{0}, I)}
\end{equation}
is the \emph{catalog Bayes factor}. Since each event is independent, we can write the catalog Bayes factor as
\begin{align}
{^{\text{(cat)}}{\mathcal{B}_{0}^{1}}} & = \frac{p(d_1, d_2, ..., d_{N_{\text{cat}}}|\mathcal{H}_{1}, I)}{p(d_1, d_2, ..., d_{N_{\text{cat}}}|\mathcal{H}_{0}, I)} \nonumber \\
& = \prod_{i = 1}^{N_{\text{cat}}} \frac{p(d_i|\mathcal{H}_{1}, I)}{p(d_i|\mathcal{H}_{0}, I)} \nonumber \\
& = \prod_{i = 1}^{N_{\text{cat}}} {^{(i)}{\mathcal{B}_{0}^{1}}},
\end{align}
where ${^{(i)}{\mathcal{B}_{0}^{1}}}$ is the Bayes factor obtained when performing the Bayesian model selection analysis on the $i$th candidate of gravitational-wave echoes event candidate. Also, we can define the \emph{catalog log Bayes factor}, which is simply
\begin{equation}\label{Eq: catalog log Bayes factor}
\ln {^{\text{(cat)}}{\mathcal{B}_{0}^{1}}} = \sum_{i = 1}^{N_{\text{cat}}} \ln {^{(i)}{\mathcal{B}_{0}^{1}}}.
\end{equation}
Hence, by multiplying the Bayes factor or adding the log Bayes factor from a catalog of gravitational-wave echoes events, we can combine the evidence of the existence of echoes in gravitational-wave data. Note that if the events share the same value of a parameter, e.g. $\Delta t_{\text{echo}}$, then the analysis is more complicated but still possible to do.

\section{Results}\label{Sec: Results}
Before performing Bayesian model selection analyses on real \acrlong{GW} events, it is necessary to validate the performance of the search methodology by performing analyses on simulated strain data first, namely strain data with Gaussian noise and an \gls{IMRE} signal of known parameters. \footnote{The signals that were manually added to the data are called injections.} By recovering the injected signal and inferring the parameters correctly, we can validate that the analysis method proposed in this \this{} will be able to find signals in real strain data. After establishing the validity of the methodology, we can sample the background and foreground distribution of the detection statistic to estimate the statistical significance of a possible gravitational-wave echoes event, and the search efficiency in simulated Gaussian noise at detectors' design sensitivities and real data in \gls{O1} of Advanced \gls{LIGO} where gravitational-wave signals were not detected.  Finally, we apply the search methodology to search for gravitational-wave echoes in \gls{O1} \gls{GW} events. The gravitational-wave strain data in \gls{O1} of Advanced \gls{LIGO} are publicly available from the Gravitational Wave Open Science Center \cite{O1DataRelease, 2015JPhCS.610a2021V}.

\subsection{Implementation}\label{Subsec: Implementation}
In this \this{}, we make use of the software package \texttt{LALSuite} developed by the \gls{LIGO} and Virgo collaborations \cite{LALSuite}. In particular, we extensively used the modules \texttt{LALSimulation} for its waveform generation interface and \texttt{LALInference} for its stochastic sampler \cite{2015PhRvD..91d2003V}. We implemented the phenomenological waveform model of gravitational-wave echoes described in Sec. \ref{Subsec: templates} in \texttt{LALSimulation}, and we have used the \gls{IMR} approximant \texttt{IMRPhenomPv2} \cite{2014PhRvL.113o1101H, 2016PhRvD..93d4006H, 2016PhRvD..93d4007K} during the echo waveform generation. We have also modified \texttt{LALInference} so that the five extra echo parameters will be sampled by the program.

It should also be noted that in theory there should be infinitely many gravitational-wave echoes. However, they are damped after each reflection from an \gls{ECO} surface and more practically we are analyzing a finite segment of gravitational-wave data, making the detection of all the echoes in an event impossible. Therefore we will only put three echoes in the template during a search, purely due to the limitation of computational power. For the purpose of model selection and statistical significance estimation, the number of echoes in a template does not matter since we are injecting \gls{IMR} signals into noise in order to estimate the background distribution of the detection statistic. It is true that putting only three echoes in a template will bias the estimation of the amplitude parameter $A$. This can be easily resolved by increasing the number of echoes in a template once we have identified an interesting \gls{GW} echoes candidate.

\subsection{Details of the validation analysis}\label{Subsec: Details of the validation analysis}
We performed our proposed search on a 8-second-long data with an \gls{IMRE} signal injected into \emph{simulated Gaussian noise} with the Advanced \gls{LIGO}-Virgo network to validate both the methodology and the implementation. We have chosen the prior distribution of the echo parameters to be uniform over a range (i.e. the \emph{prior range}), and that the echoes will not overlap in time domain. The prior ranges of the parameters used are listed in Table \ref{Table: PE Prior}. The prior range for $A$ was chosen as such because we do not expect the amplitude of echoes to be greater than the amplitude of the \acrlong{IMR} part of a signal. However, this is not a stringent requirement and can be easily relaxed. As for the prior range for $\gamma$, it was chosen as such because we expect echoes to be damped after each reflection from an \gls{ECO} surface. The prior range for $t_{0}$ was chosen such that we are truncating approximately the ringdown part of a signal. As for the prior ranges for both $t_{\text{echo}}$ and $\Delta t_{\text{echo}}$, they depend on our knowledge of the position of the surface of an \gls{ECO}. For the purpose of this work, they were chosen to be wide enough such that their predicted values for all the \gls{GW} events detected in \gls{O1} as calculated in Ref. \cite{2017PhRvD..96h2004A} fall within their corresponding prior ranges, which are sensible.

The \gls{IMR} parameters of \emph{this particular injected signal}, such as masses and spins, were chosen to be close to the inferred values of GW150914 \cite{2016PhRvL.116x1102A} and the injected echo parameters were chosen \emph{randomly} over the prior range. It should be noted that in this work \emph{we do not assume \gls{IMR} parameters were known a priori and they were allowed to vary during the validation analysis together with the echo parameters}. This is because the \gls{IMR} parameters would affect the determination of the echo waveform used and thus the uncertainties in inferring \gls{IMR} parameters would also propagate to the the search for echoes. This particular injection has a log Bayes factor of $11.5$, and a network optimal SNR of $63.8$, which will be a realistic value when Advanced LIGO-Virgo detectors reach their design sensitivities.

\subsection{Parameter estimation}\label{Subsec: PE on simulated data}
\begin{figure*}[!pht]
\begin{center}
\includegraphics[width=1.0\textwidth]{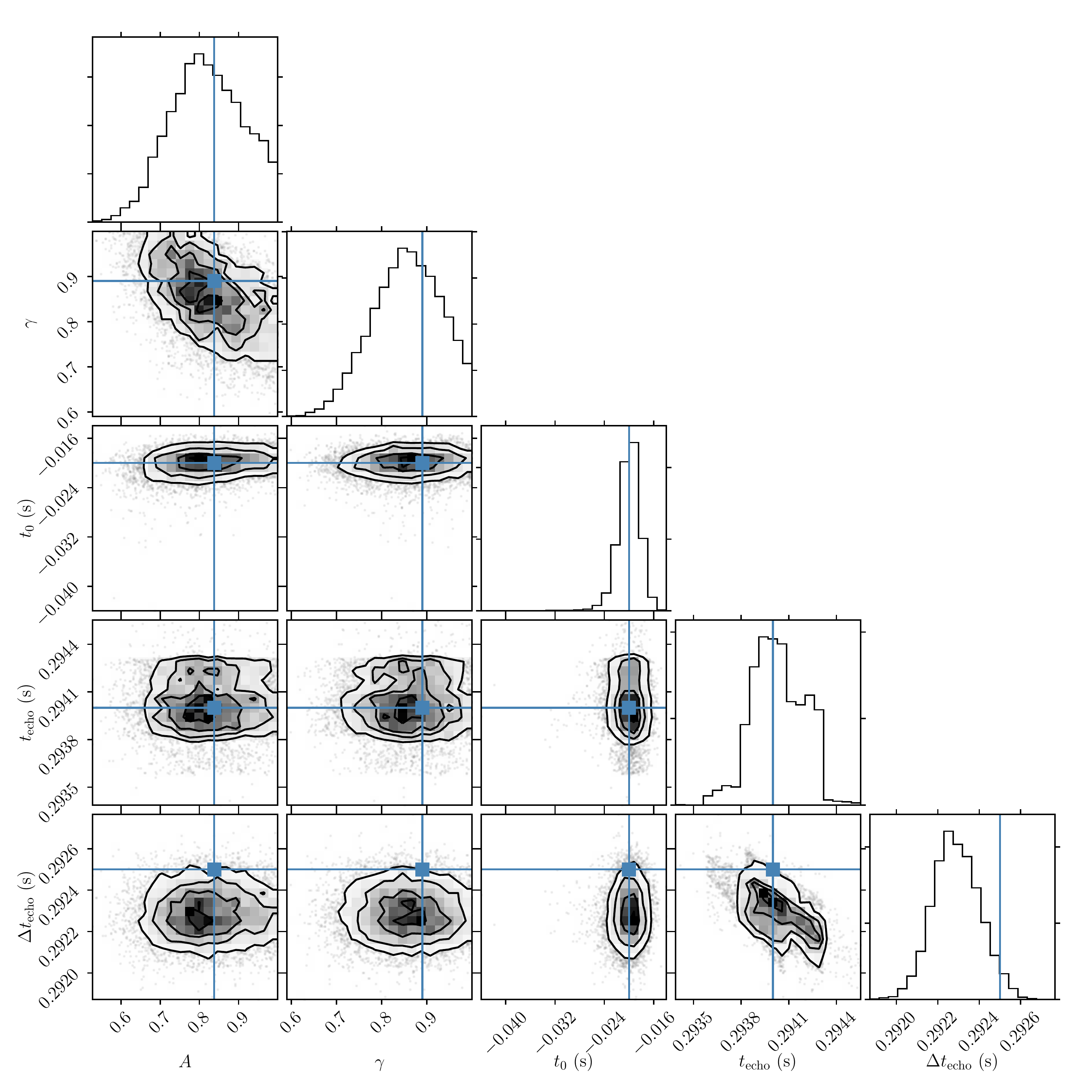}
\caption{\label{Fig: Corner}A corner plot of the posterior samples from the parameter estimation on simulated data as described in Secs. \ref{Subsec: Details of the validation analysis} and \ref{Subsec: PE on simulated data}. If we are to use the proposed search methodology to search for gravitational-wave echoes in real data, then the search needs to accurately recover the injected \gls{IMRE} signal. The blue solid lines represent the injected values for each parameter. Along the diagonal are histograms of the estimated 1D marginal posterior probability distribution for each parameter. The histograms show that the recovered parameters are both accurate (close to the injected value) and precise (narrow posterior distribution). For example, the 1D marginal posterior probability density of $\Delta t_{\text{echo}}$ is very narrow compared to its prior range tabulated in Table \ref{Table: PE Prior}, and the peak of the posterior probability distribution is very close to the injected value. The off-diagonal plots are the two-dimensional histograms of the estimated joint posterior probability distribution of each pair of parameters, which show the correlation between pairs of parameters. We conclude that the search methodology is able to infer the values of the echo parameters in actual analyses on candidate gravitational-wave echoes events as it has successfully recovered the injected \gls{IMRE} signal in the validation analysis accurately and precisely.}
\end{center}
\end{figure*}

As an output of our search methodology, the set of posterior samples allows us to perform the parameter estimation on the simulated data. The search needs to accurately recover the injected \gls{IMRE} signal if we are to use the proposed search methodology to search for gravitational-wave echoes in real data. 
A visualization of the sampled posterior distributions, i.e. a corner plot, that shows the estimated one-dimensional (1D) marginal posterior probability distribution for each parameter and joint posterior probability distribution for each pair of parameters, is shown in Fig. \ref{Fig: Corner}. We see that the inferred values of the echo parameters are both accurate (close to the injected value) and precise (narrow posterior distribution), especially for the time-related parameters. For example, we see from the 1D histogram of $\Delta t_{\text{echo}}$ in Fig. \ref{Fig: Corner} that the \gls{MAP} is very close to the injected value (represented by the vertical blue solid line), and the 90\% Bayesian credible interval ([0.2921, 0.2925] s) is much narrower than the prior range ([0.05, 0.5] s), which means that the range is shrunk by about 99.91\%.

\begin{table}[h!]
\begin{center}
\begin{ruledtabular}
\begin{tabular}{lD{,}{,}{3.5}c}
\multicolumn{1}{c}{Parameter} & \multicolumn{1}{c}{Prior range} \\ \hline
$A$ & [0.0, 1.0] \\
$\gamma$ & [0.0, 1.0] \\
$t_{0}$ (s) & [-0.1, 0.01] \\
$t_{\text{echo}}$ (s) & [0.05, 0.5] \\
$\Delta t_{\text{echo}}$ (s) & [0.05, 0.5]          
\end{tabular}
\end{ruledtabular}
\caption{\label{Table: PE Prior}The prior range of the echo parameters. The prior distribution of each parameter is \emph{uniform} over the respective prior range. Refer to the main text for the justification for the choice of prior ranges.}
\end{center}
\end{table}

As for the amplitude-related parameters $A$ and $\gamma$, the parameter estimation is not as accurate and precise as for the time-related parameters. For instance, we see that the range for $A$ does not shrink as much compared with $\Delta t_{\text{echo}}$ (only by about 60\%).
This is not surprising because the time-related parameters can be inferred using the coherence of the strain with a template, while the amplitude-related parameters can only be inferred using noisy strain data.

By examining the corner plot for recovered \gls{IMR} parameters (not shown here), we conclude that the parameter estimation of \gls{IMR} parameters was not significantly affected by the introduction of five extra parameters.

Therefore, from the parameter estimation, we conclude that we have correctly implemented the gravitational-wave echoes waveform model and modified the sampler, and more importantly the search methodology is able to infer the values of echo parameters in actual analyses on candidate gravitational-wave echoes events as it has successfully recovered the injected \gls{IMRE} signal accurately and precisely in this validation analysis.

\subsection{Model selection}\label{Model selection on simulated data}
\subsubsection{Statistical significance estimation of a candidate gravitational-wave echoes event}\label{Subsec: Background estimation}
To estimate the statistical significance of a gravitational-wave echoes candidate, we sampled the null distribution $p(\ln \mathcal{B}|\mathcal{H}_{0})$ of the detection statistic by performing background runs, i.e. data \emph{with an \gls{IMR} signal injected} (so that the null hypothesis $\mathcal{H}_{0}$ is true).
The \gls{IMR} parameters of the injection set used to estimate the background distribution were chosen to be representative of what Advanced LIGO and Advanced Virgo would detect, and \emph{were not fixed to be the same as a particular gravitational-wave event}. We will discuss this choice in Sec. \ref{Subsec: Discussion on injection sets}.

\begin{figure*}[ht!]
\begin{center}
\subfigure{\includegraphics[width=1.0\columnwidth]{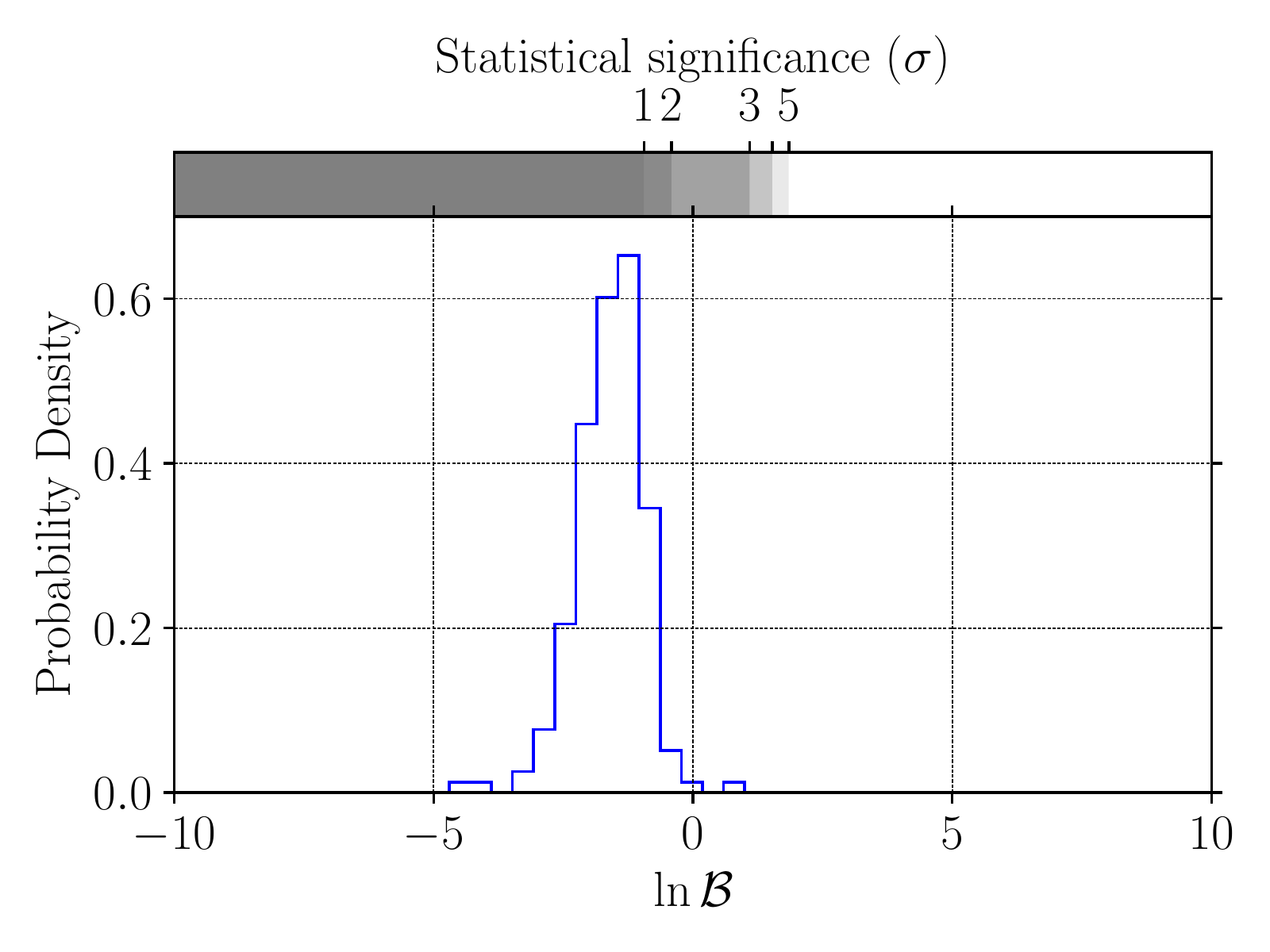}}
\subfigure{\includegraphics[width=1.0\columnwidth]{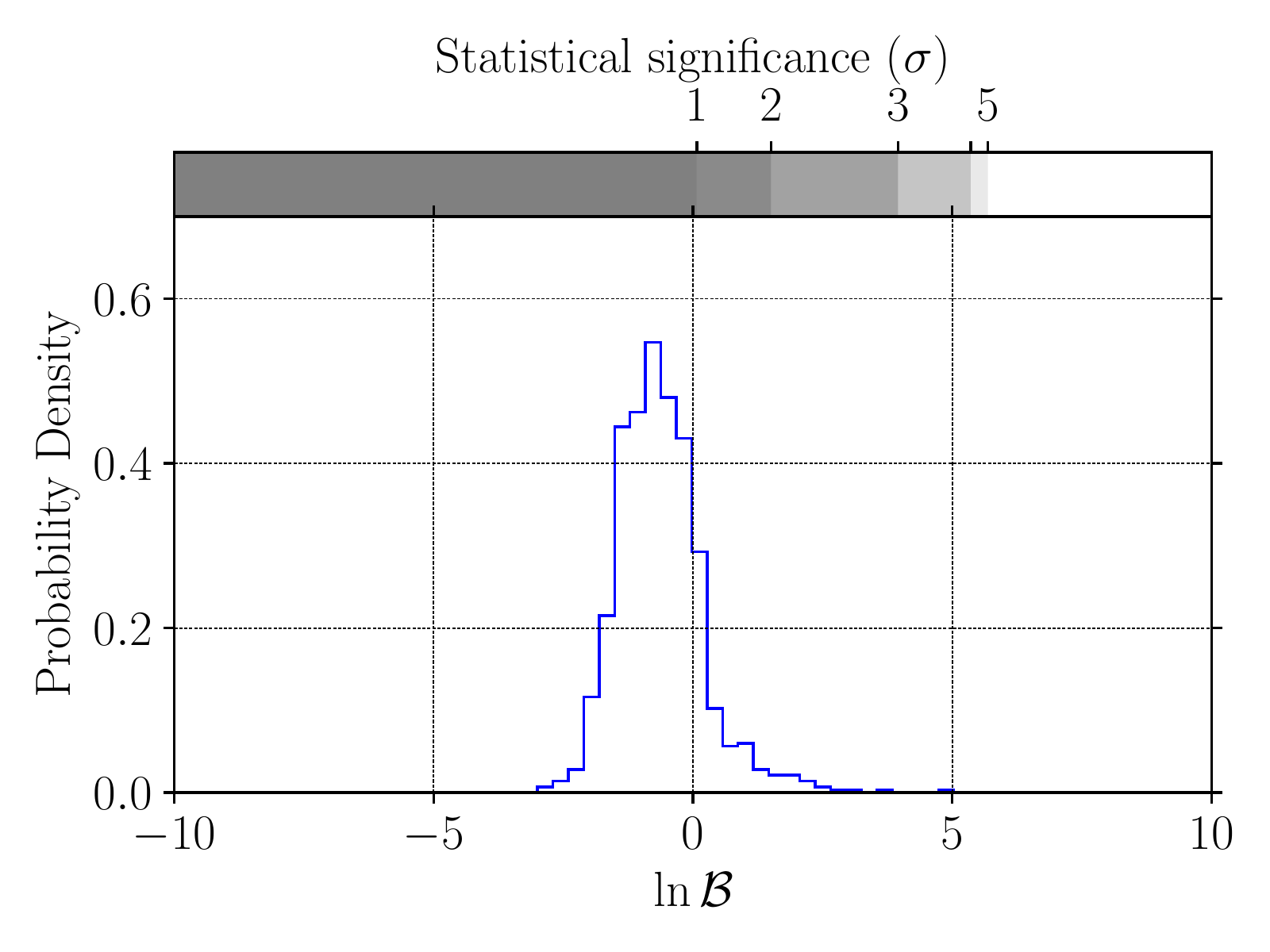}}
\caption{\label{Fig: Null distribution}To estimate the statistical significance of a potential gravitational-wave echoes event for simulated Gaussian noise and real noise during \gls{O1} of Advanced \gls{LIGO}, we sampled the null distribution $p(\ln \mathcal{B}|\mathcal{H}_{0})$ of the detection statistic by performing background runs, i.e. data with an \gls{IMR} signal injected. The histograms of null distribution for the case of Gaussian noise with $192$ samples and for the case of \gls{O1} noise with $953$ samples are plotted in the left and right panels respectively. The gray-scale bar in the top panel shows the statistical significance corresponding to the detection statistic (extrapolating from the $3\sigma$ region to the $5\sigma$ region). \textit{Left panel:} For the case of simulated Gaussian noise, we see that the null distribution peaks at about $\ln \mathcal{B} \approx -1$, and the tail of the distribution extends only slightly towards $\ln \mathcal{B} > 0$. This means that it is unlikely for Gaussian noise to mimic gravitational-wave echoes. \textit{Right panel:} For the case of real noise during \gls{O1}, we see that the distribution also peaks roughly at $\ln \mathcal{B} \approx -1$. However, the noise extends the tail of the distribution more significantly than in the case of Gaussian noise. This means that it is likely for real detector noises to mimic the effects of gravitational-wave echoes.}
\end{center}
\end{figure*}

The histograms of the sampled null distribution of the \emph{individual log Bayes factor} for simulated Gaussian noise (with $192$ samples) and real noise during \gls{O1} of Advanced \gls{LIGO} (with $953$ samples) \footnote{The number of samples for the background distribution in simulated Gaussian noise is less than that for real \gls{O1} noise because of the lack of computational resources} are shown in the left and right panels of Fig. \ref{Fig: Null distribution} respectively. The gray-scale bar in the top panel shows the statistical significance corresponding to the detection statistic. It should be noted that the $p$-value was obtained by extrapolation for the $\gtrsim 3\sigma$ region, as sampling the $> 5\sigma$ region would require roughly $10^7$ samples. For the case of simulated Gaussian noise in the left panel, we see that the null distribution peaks at about $\ln \mathcal{B} \approx -1$, and the tail of the distribution extends only slightly to $\ln \mathcal{B} > 0$. This means that it is unlikely for Gaussian noise to mimic gravitational-wave echoes. For the case of real noise during \gls{O1} in the right panel, we see that the distribution also peaks roughly at $\ln \mathcal{B} \approx -1$. However, the noise extends the tail of the distribution more significantly than in the case of Gaussian noise. This means that it is more likely for real detector noises to mimic the effects of gravitational-wave echoes.

In particular, we injected an \gls{IMRE} injection with echo parameters that \emph{Abedi et al.~claimed to have found in GW150914} into simulated Gaussian noise with the Advanced \gls{LIGO}-Virgo network, and the detection statistic was found to be
\[ \ln \mathcal{B}_{\text{detected, Gaussian}} = -0.2576 < 0. \]
This means that in the Bayesian point of view, the data slightly favor the null hypothesis that the data do not contain echoes.
We compute the $p$-value and the corresponding statistical significance, given that the noise is Gaussian, as follows:
\begin{align*}
p\text{-value} & = 0.01275, \\
\text{statistical significance} & = 2.234 \sigma .\\
\end{align*}
This suggests that what was claimed to be found by Abedi et al.~in GW150914, even for gravitational-wave detectors operating at design sensitivities and Gaussian noise, does not have sufficient statistical significance to claim a detection (i.e. $\geq 5 \sigma$) in the frequentist approach, and it is also inconclusive whether the data favor the existence of echoes in the data in the Bayesian approach. From the fact that in Fig. \ref{Fig: Null distribution} the null distribution for \gls{O1} real noise is more skewed to the right, we can expect that the statistical significance of what Abedi et al.~ had found is small and consistent with noise.

Table \ref{Table: detection statistic threshold} tabulates the values of the detection statistic $\ln \mathcal{B}$ that correspond to different levels of statistical significance in Gaussian and \gls{O1} backgrounds. If we want to make a gold-plated detection of gravitational-wave echoes, i.e. having statistical significance $\geq 5 \sigma$, we can set the detection threshold as
\begin{align*}
\ln \mathcal{B}_{\text{threshold, Gaussian}} & = \GaussianThreshold{}, \\
\ln \mathcal{B}_{\text{threshold, O1}} & = \OOneThreshold{},
\end{align*}
in the case of Gaussian noise and real \gls{O1} noise respectively, so that any gravitational-wave echoes detection with a detection statistic greater than or equal to this threshold is a $\geq 5\sigma$ detection of echoes.

\begin{table}[h!]
\begin{center}
\begin{ruledtabular}
\begin{tabular}{cd{3.5}d{3.5}}
\multicolumn{1}{p{0.25\columnwidth}}{\centering{Statistical significance}} & \multicolumn{1}{p{0.25\columnwidth}}{\centering{Detection statistic \emph{(Gaussian noise)}}} & \multicolumn{1}{p{0.25\columnwidth}}{\centering{Detection statistic\\ \emph{(O1 noise)}}} \\ \hline
$1 \sigma$ & \GaussianOneSigma{} & \OOneOneSigma{} \\
$2 \sigma$ & \GaussianTwoSigma{} & \OOneTwoSigma{} \\
$3 \sigma$ & \GaussianThreeSigma{} & \OOneThreeSigma{} \\
$4 \sigma$ & \GaussianFourSigma{} & \OOneFourSigma{} \\
$5 \sigma$ & \GaussianFiveSigma{} & \OOneFiveSigma{} \\       
\end{tabular}
\end{ruledtabular}
\caption{\label{Table: detection statistic threshold}The values of the detection statistic $\ln \mathcal{B}$ and its corresponding statistical significances in both Gaussian and \gls{O1} backgrounds. If we want to make a gold-plated detection of gravitational-wave echoes, i.e. with a statistical significance $\geq 5 \sigma$, we can set the detection statistic threshold as $\ln \mathcal{B}_{\text{threshold, Gaussian}} = \GaussianThreshold{}$ in the case of Gaussian noise, and $\ln \mathcal{B}_{\text{threshold, O1}} = \OOneThreshold{}$ so that any gravitational-wave echoes detection with a detection statistic greater than or equal to this threshold is gold-plated.}
\end{center}
\end{table}

\subsection{Search sensitivity, efficiency and accuracy}\label{Subsec: search sensitivity, efficiency and accuracy of simulated data}

\subsubsection{Sensitive parameter space of the search}\label{Subsubsec: sensitive parameter space}
Given the detection statistic threshold $\ln \mathcal{B}_{\text{threshold}}$, we would like to know what part of the parameter space of echoes we are able to see in the \emph{optimal case}, where the gravitational-wave detectors are operating at design sensitivities and the instrumental noise is Gaussian. To achieve this, we performed analyses on simulated data injected with an IMRE signal with different values of the echo parameters of interest. In this particular study, the \gls{IMR} parameters were fixed to be \emph{GW150914-like}. We will be focusing on the two parameters that are of the most astrophysical interest: the time interval between echoes $\Delta t_{\text{echo}}$ and the relative amplitude $A$. The two parameters were varied one at a time.

\begin{figure*}[ht!]
\subfigure{\includegraphics[width=1.0\columnwidth]{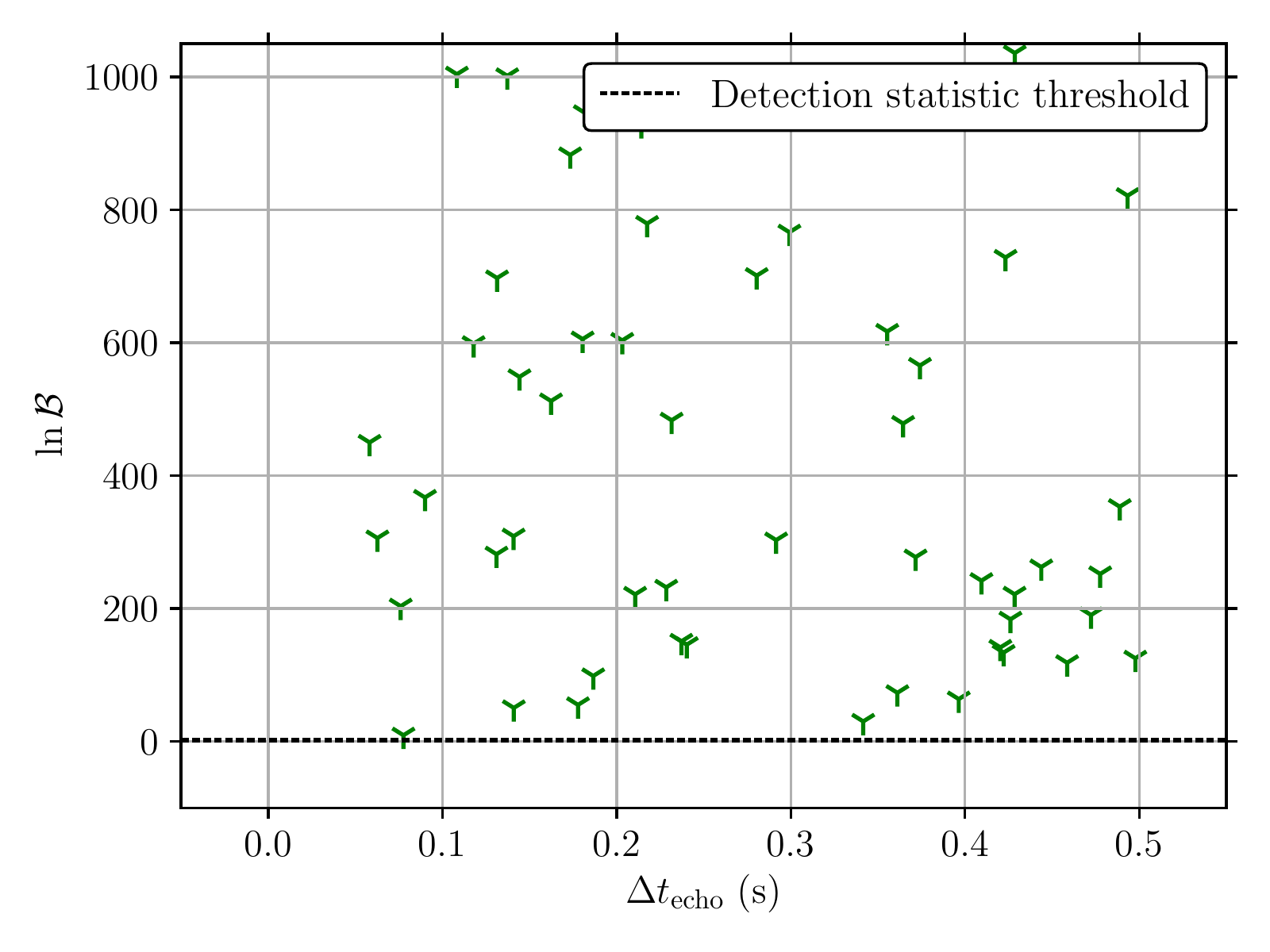}}
\subfigure{\includegraphics[width=1.0\columnwidth]{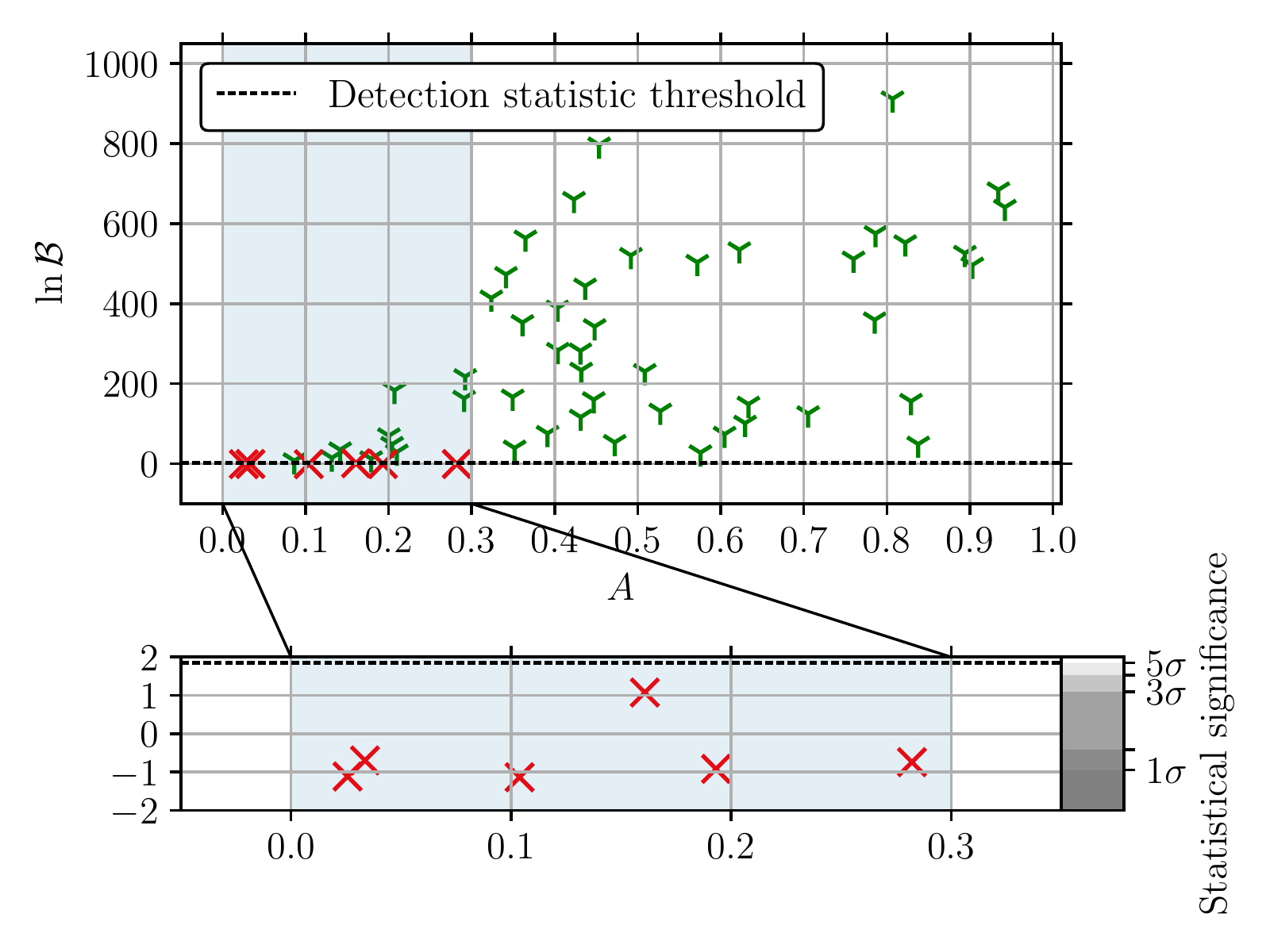}}
\caption{\label{Fig: Sensitive parameter space}To investigate which part of the parameter space of echoes we are able to see in the optimal case, namely the gravitational-wave detectors are operating at their design sensitivities and the instrumental noise is Gaussian, we performed analyses on simulated data injected with an IMRE signal with different values of the echo parameters of interest. \textit{Left panel}: A plot of the detection statistic $\ln \mathcal{B}$ as a function of $\Delta t_{\text{echo}}$, with other echo parameters fixed to $(A = 0.6, \gamma = 0.89, t_{0} = -0.02 \text{\,s}, t_{\text{echo}} = 0.2940 \text{\,s})$. \textit{Right panel}: A plot of the detection statistic $\ln \mathcal{B}$ as a function of $A$, with the other echo parameters fixed to $(\gamma = 0.89, t_{0} = -0.02 \text{\,s}, t_{\text{echo}} = 0.2940 \text{\,s}, \Delta t_{\text{echo}} = 0.2925 \text{\,s})$. The horizontal dashed line in each plot corresponds to the detection statistic threshold of $5 \sigma$ significance. Injections with a detection statistic exceeding or equal to the threshold are marked with a green `\texttt{Y}', whereas injections with detection statistic lower than the threshold are marked with a red `\texttt{X}'. From the left panel, we see that there is \textit{no trend} for how the detection statistic is distributed with different values of $\Delta t_{\text{echo}}$, and that the search is able to detect gravitational-wave echoes with a range of $\Delta t_{\text{echo}}$ (more specifically $[0.05, 0.5]$ s) as expected since different values of $\Delta t_{\text{echo}}$ only shift the echoes in time, and whether the search is able to find echoes or not should not depend on their time of occurrence as long as they do not overlap. Therefore, fixing the values of the time-related echo parameters when investigating the sensitive parameter space of $A$ is justified. As for the relative amplitude $A$, we see from the right panel that there is a trend that signals with \textit{smaller values of $A$ have smaller values of the detection statistic}, and that the search can only pick up echoes with $A \gtrsim 0.3$ with $\geq 5 \sigma$ significance. This is expected because the amplitude of echoes is damped and echoes with small amplitudes are buried in noise. It should be noted that we found some injections with echo amplitudes $A \gtrsim 0.1$, and the number we quoted for the sensitive parameter space for $A$ is based on the loudest echo injections that were missed.}
\end{figure*}

Figure \ref{Fig: Sensitive parameter space} shows plots of the detection statistic $\ln \mathcal{B}$ as a function of $\Delta t_{\text{echo}}$ with the other echo parameters fixed to $(A = 0.6, \gamma = 0.89, t_{0} = -0.02 \text{\,s}, t_{\text{echo}} = 0.2940 \text{\,s})$(left) and as a function of $A$ with other echo parameters fixed to $(\gamma = 0.89, t_{0} = -0.02 \text{\,s}, t_{\text{echo}} = 0.2940 \text{\,s}, \Delta t_{\text{echo}} = 0.2925 \text{\,s})$ (right) respectively. The horizontal dashed line in each plot corresponds to the detection statistic threshold of $5 \sigma$ significance. Injections with a detection statistic exceeding or equal to the threshold are marked with a green `\texttt{Y}', whereas injections with detection statistic lower than the threshold are marked with a red `\texttt{X}'. From the left panel, we see that there is \textit{no trend} for how the detection statistic is distributed with different values of $\Delta t_{\text{echo}}$, and that the search is able to detect gravitational-wave echoes with a range of $\Delta t_{\text{echo}}$ (more specifically $[0.05, 0.5]$ s) as expected since different values of $\Delta t_{\text{echo}}$ only shift the echoes in time, and whether the search is able to find echoes or not should not depend on their time of occurrence as long as they do not overlap. Therefore, fixing the values of time-related echo parameters when investigating the sensitive parameter space of $A$ is justified. As for the relative amplitude $A$, we see from the right panel that there is a trend that signals with \textit{smaller values of $A$ have smaller values of the detection statistic}, and that the search can only pick up echoes with $A \gtrsim 0.3$ with $\geq 5 \sigma$ significance. This is expected because the amplitude of echoes is damped and echoes with small amplitudes are buried in noise. This finding is consistent with that of Westerweck \etal{} that only injections with a strain amplitude $\gtrsim 10^{-22}$ in the echoes part could have the amplitude parameter $A$ recovered accurately \cite{2018PhRvD..97l4037W}. In Ref. \cite{2018arXiv181104904N}, a plot similar to the right panel of Fig. \ref{Fig: Sensitive parameter space} was also shown \cite{2018arXiv181104904N}, but it was unclear in their paper that at what value of their detection statistic (log Bayes factor for signal versus Gaussian noise, which is \emph{different} from what we adopted in this \this{}) they are claiming a significant detection of echoes, and we will differ the discussion of the differences between two approaches in Sec. \ref{Subsubsec: AEI Inference}. It should also be noted that there are injections with echo amplitudes $A \gtrsim 0.1$ that are found by our search, and the number we quoted for the sensitive parameter space for $A$ is based on the loudest echo injection that were missed.

\subsubsection{Foreground distribution and search efficiency}\label{Subsubsec: Foreground distribution and search efficiency}
To compute the efficiency $\zeta$ of the search as described in Sec. \ref{Subsubsec: Search efficiency}, the foreground distribution $p(\ln \mathcal{B}|\mathcal{H}_{1})$ of the detection statistic was sampled by performing foreground runs, i.e. analyses on simulated data with \gls{IMRE} signals injected. The \gls{IMR} parameters of the injection set used to estimate the foreground distribution were chosen to be representative of what Advanced LIGO and Advanced Virgo would detect, and \emph{were not fixed to be the same as a particular gravitational-wave event}. We will discuss this choice in Sec. \ref{Subsec: Discussion on injection sets}. As for the five echo parameters, their values were drawn randomly from the same distributions described in Table \ref{Table: PE Prior}.

A numerical integration of Eq. \ref{Eq: Efficiency} on the foreground distribution sampled gives the efficiency of the search as
\begin{align*}
 \zeta_{\text{Gaussian}} & = \GaussianEfficiencyDecimal{}, \\
 \zeta_{\text{O1}} & = \OOneEfficiencyDecimal{}.
\end{align*}
That means, in the frequentist language, that the search has a probability of \GaussianEfficiencyDecimal{} in the case of Gaussian noise and \OOneEfficiencyDecimal{} in the case of \gls{O1} of detecting the existence of echoes \emph{marginalized over a set of IMR and echo parameters} (with echo parameters drawn uniformly from the priors in Table \ref{Table: PE Prior}), given that the data contain gravitational-wave echoes and that the detection has a $\geq 5 \sigma$ significance.

\subsubsection{Search accuracy}
\begin{figure}[h!]
\begin{center}
\includegraphics[width=1.0\columnwidth]{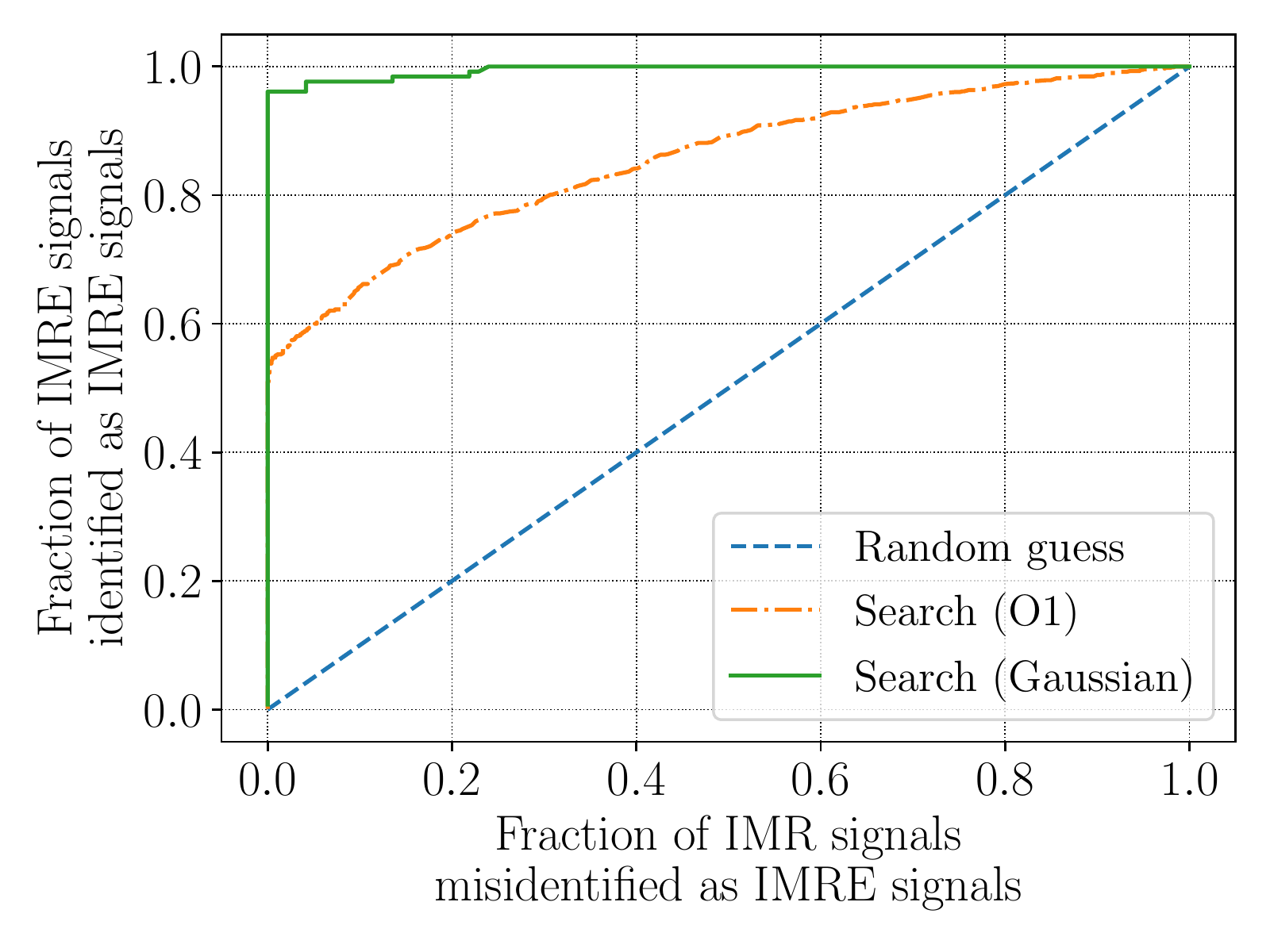}
\caption{\label{Fig: ROC Curve}To gauge the performance of our proposed search methodology in terms of the ability to classify IMR and IMRE signals, the \gls{ROC} curves for searches in O1 noise (orange dash-dotted line) and Gaussian simulated noise (green solid line) respectively are shown. We see that the search in simulated Gaussian noise performs better than the search in O1 noise as expected because the fraction of IMRE signals identified as IMRE signals (with echo parameters drawn uniformly from the priors in Table \ref{Table: PE Prior}) for the simulated Gaussian noise case is higher than that for the O1 noise case for a given fraction of IMR signals misidentified as IMRE signals. Also, both searches in simulated Gaussian noise and O1 noise outperform the random guess (blue dashed line) for the same reason described above.}
\end{center}
\end{figure}

Given the detection statistic threshold $\ln \mathcal{B}_{\text{threshold, Gaussian}} = \GaussianThreshold{}$ for the case of simulated Gaussian noise and $\ln \mathcal{B}_{\text{threshold, O1}} = \OOneThreshold{}$ for the case of \gls{O1}, we make the claim that the data contain echoes only when the detection statistic is greater than or equal to the threshold. To gauge the performance of our proposed search methodology in terms of the ability to classify \gls{IMR} and \gls{IMRE} signals, a plot of the \gls{ROC} curve is shown in Fig. \ref{Fig: ROC Curve} for both the simulated Gaussian noise case and the \gls{O1} noise case. The \gls{ROC} curve shows the fraction of \gls{IMRE} signals that the search has properly identified as \gls{IMRE} signals (also known as the \emph{true positive rate} for binary classifiers) given the fraction of \gls{IMR} signals that the search has incorrectly identified as \gls{IMRE} signals (also known as the \emph{false positive rate}). Equivalently, it can also be interpreted as showing how the efficiency of a search changes with the detection statistic threshold.
A more sensitive search will have a higher true positive rate for a given false positive rate. For a random guess, the true positive rate is the same as the false positive rate. Therefore, the \gls{ROC} curve of a useful search should be to the left of the diagonal \gls{ROC} curve. From Fig. \ref{Fig: ROC Curve}, we see that the search in simulated Gaussian noise performs better than the search in \gls{O1} noise as expected, and both of the searches perform better than random guessing.

\subsection{Demonstration of combining evidence from multiple gravitational-wave echoes events}\label{Subsec: Demonstration of combining evidence}
From the investigation of the sensitive parameter space of the search in the optimal case as described in Sec. \ref{Subsubsec: sensitive parameter space}, we see that the gravitational-wave echoes will need to have a relative strength $A \gtrsim 0.3$ in order to be picked up by our search with $\geq 5\sigma$ significance in the optimal Gaussian noise case. Statistically we can incorporate results from multiple events so that the detection statistic of many weak signals can be added positively to stand out from the combined background, which adds negatively. Instead of combining the posterior distribution of echo parameters, which assumed the echo parameters for each event to be the same, we add the detection statistic log Bayes factor to form the \emph{catalog log Bayes factor}, which is described in Sec. \ref{Subsec: Combine evidence}, and this approach only assumes each event to be independent and does not require the parameters of the events to be identical; indeed, it is assumed that they are all different for each event.

\begin{figure*}[ht!]
\subfigure{\includegraphics[width=1.0\columnwidth]{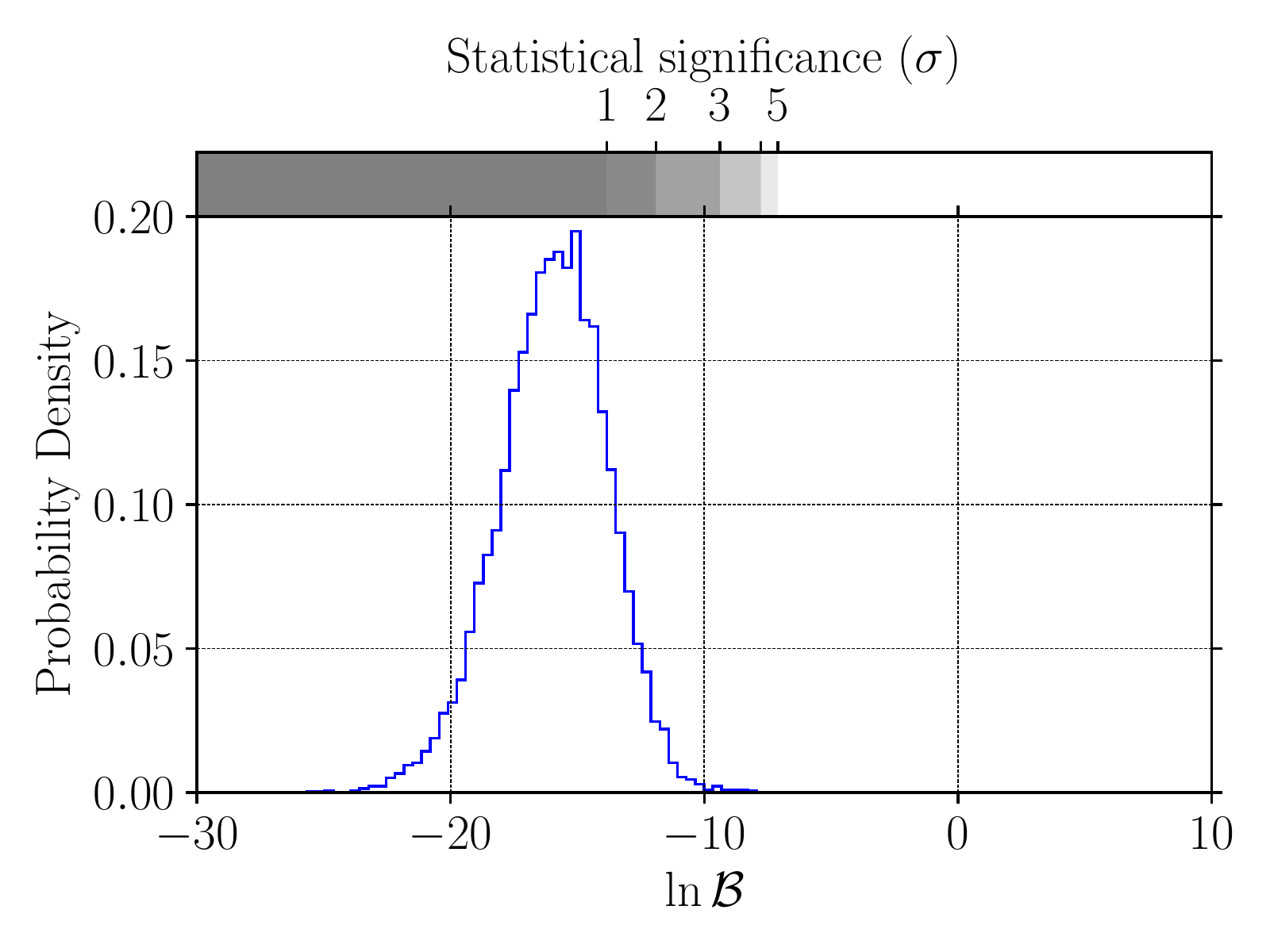}}
\subfigure{\includegraphics[width=1.0\columnwidth]{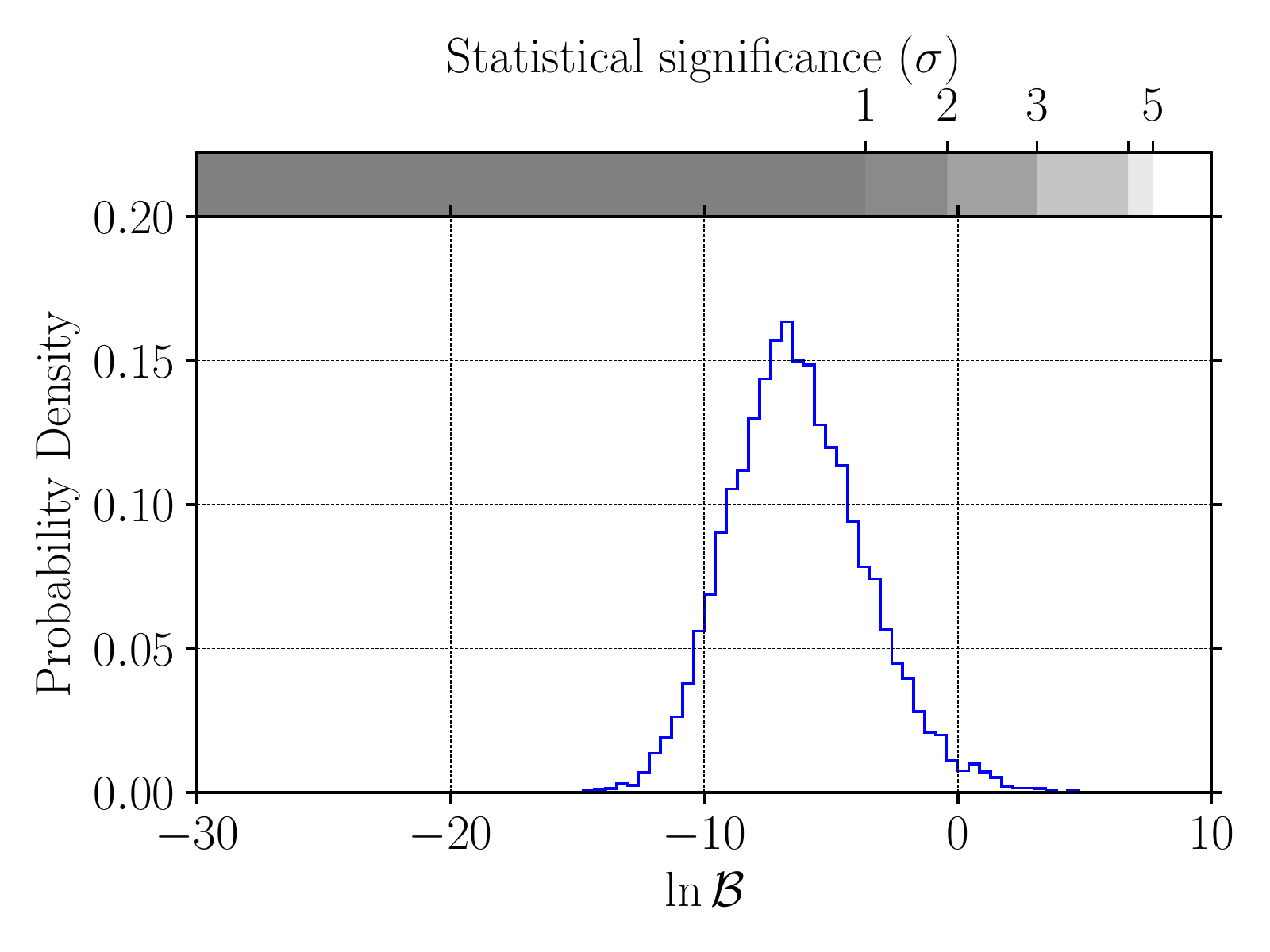}}
\caption{\label{Fig: Catalog}To detect gravitational-wave echoes which may be individually too weak to be detected, we can instead make a detection statement on \emph{a collection of events} (i.e. a catalog). The histograms of the sampled null distribution for the \emph{catalog log Bayes factor} for simulated Gaussian noise and real noise during \gls{O1} of Advanced \gls{LIGO} (both with $10000$ samples) with a catalog size $N_{\text{cat}} = 10$ are shown in the left and right panels respectively. The gray-scale bar in the top panel shows the statistical significance corresponding to the detection statistic. Compared to the histograms for individual log Bayes factors shown in Fig. \ref{Fig: Null distribution}, we see that the peak of the null distribution for the catalog log Bayes factor for both the case of Gaussian noise and \gls{O1} noise is shifted to be more negative as expected. Since weak signals add positively while the background events add negatively in the catalog log Bayes factor, weak gravitational-wave echoes signals can be detected as a whole, and a statement about whether there are echoes in a collection of potential gravitational-wave echoes candidates can be made but not about which individual event/events in the catalog has/have echoes.}
\end{figure*}

Here we demonstrate how combining multiple events can help us to detect weak echoes and make a detection statement about \emph{a collection of events} (i.e. a catalog). Suppose we have ten potential gravitational-wave echoes events, which is to say we have $N_{\text{cat}} = 10$ events in the catalog. The catalog log Bayes factor is simply the sum of the log Bayes factors of each individual event according to Eq. \ref{Eq: catalog log Bayes factor}. We sampled the null distribution for the catalog log Bayes factor 
$p(\ln ^{\text{(cat)}} \mathcal{B}|\mathcal{H}_{0}) = p(\sum_{i = 1}^{N_{\text{cat}}} \ln {^{(i)}{\mathcal{B}_{0}^{1}}} |\mathcal{H}_{0})$ by picking $N_{\text{cat}} = 10$ events from the background distribution and computing the corresponding catalog log Bayes factor. The histograms of the sampled null distribution for the \emph{catalog log Bayes factor} for simulated Gaussian noise and real noise during \gls{O1} of Advanced \gls{LIGO} (both with $10000$ samples in the sampled background distributions) are shown in the left and right panels of Fig. \ref{Fig: Catalog} respectively. The gray-scale bar in the top panel shows the statistical significance corresponding to the detection statistic. Compared to the histograms for individual log Bayes factors shown in Fig. \ref{Fig: Null distribution}, we see that the peak of the null distribution for the catalog log Bayes factor for both the case of Gaussian noise and \gls{O1} noise is shifted to be more negative as expected. If we focus on the case of Gaussian noise, and assume that the mean of the log Bayes factor $\langle \ln \mathcal{B} \rangle$ is roughly $-1$, then the mean of the catalog log Bayes factor of the size of $N_{\text{cat}} = 10$ should be 
\[
\langle \ln ^{\text{(cat)}}\mathcal{B} \rangle \approx N_{\text{cat}} \langle \ln \mathcal{B} \rangle = -10,
\]
which is indeed the case in the right panel of Fig. \ref{Fig: Catalog}. If we assume, for the sake of demonstration, that we have observed ten gravitational-wave echoes events similar to what Abedi et al. claimed to have found in GW150914, namely the individual log Bayes factor is about $-0.2576$, then the catalog log Bayes factor will then become roughly
\[
\ln ^{\text{(cat)}}\mathcal{B} \approx N_{\text{cat}} \times -0.2576 = -2.576,
\]
and thus we can make a statement with $\geq 5\sigma$ significance that there are gravitational-wave echoes in one or more events in the catalog, but we will not be able to pinpoint which event has echoes. 

From this example, we see that by combining the Bayesian evidence from multiple events, we can statistically make a detection statement about whether there are echoes in a collection of potential gravitational-wave echoes candidates, which may be too weak to be detected individually.
 
\subsection{Search results for Advanced LIGO's first observing run data}\label{Subsec: O1 Search Results}
We applied the search methodology described above to search for gravitational-wave echoes in Advanced LIGO's \gls{O1} data. The prior distribution of the five echo parameters were chosen to be uniform over the respective prior ranges tabulated in Table \ref{Table: PE Prior}. As for the \gls{IMR} parameters, they were allowed to vary during the analyses and were not fixed to any particular values. For GW150914 and GW151012, we used $8$-second-long data with three echoes in the \gls{IMRE} waveforms for parameter estimation. As for GW151226, we used $16$-second-long data with \emph{ten echoes} in the \gls{IMRE} waveform. The difference in the length of data used is because the duration (starting from $30$ Hz to merger) of the signal for GW151226 is slightly more than $1.5$ s \cite{2016PhRvX...6d1015A}, while those for GW150914 and GW151012 are roughly less than $0.5$ s \cite{2016PhRvX...6d1015A}. This is due to the fact that GW151226 has a lower chirp mass compared to the other two events. Hence, we used a longer data segment for the analysis of GW151226, which enabled us to include more echoes in the \gls{IMRE} waveforms.

\begin{figure}[h]
\includegraphics[width=1.0\columnwidth]{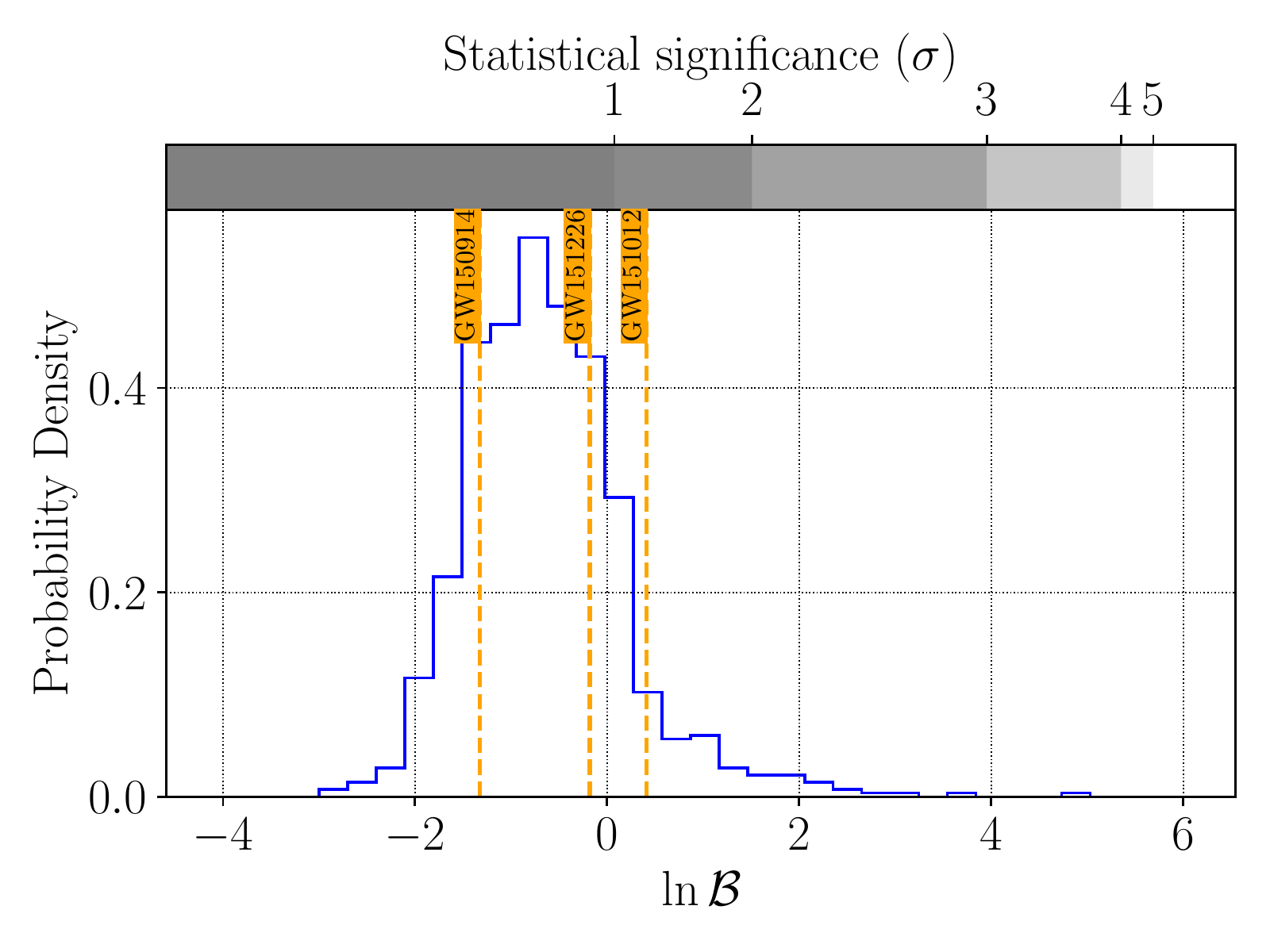}
\caption{\label{Fig: O1 results}The histogram of the sampled null distribution for the \emph{log Bayes factor} $\ln \mathcal{B}$ for Advanced LIGO's \gls{O1} data. This is the same plot as in the right panel of Fig. \ref{Fig: Null distribution} with the detected values of the log Bayes factor for the three events in \gls{O1} (namely GW150914, GW151012 and GW151226). For GW150914 and GW151226, their detection statistics are less than zero, with a statistical significance less than $1\sigma$. As for GW151012, although the detection statistic for GW151012 is slightly greater than zero, it is still well within the background distribution. In fact, the statistical significance is only about $1.4\sigma$. Therefore, we conclude that no significant evidence was found to support the detection of gravitational-wave echoes in \gls{O1} data.}
\end{figure}

Table \ref{Table: detection statistics for O1} tabulates the detection statistic and the corresponding statistical significance and $p$-value for the three events (GW150914 \cite{PhysRevLett.116.061102,2016PhRvX...6d1015A}, GW151012 \cite{2016PhRvX...6d1015A}, and GW151226 \cite{PhysRevLett.116.241103, 2016PhRvX...6d1015A}) in \gls{O1}. None of the events have a detection statistic greater than the $5\sigma$-detection threshold. Although the detection statistics used in the analyses were different, the ordering of the events by their statistical significance is consistent with that reported by Nielsen \etal{}\cite{2018arXiv181104904N}. For GW150914, the detection statistic is indeed less than the \emph{upper bound} estimated in Sec. \ref{Subsec: Background estimation} for the case of the signal that Abedi et al. claimed to have found in GW150914 \cite{2017PhRvD..96h2004A} in simulated Gaussian noise. In particular, GW151012 has the highest detection statistic among the three, and the value is greater than zero. However, that value is well within the background distribution we showed in the right panel of Fig. \ref{Fig: Null distribution}. Figure \ref{Fig: O1 results} summarizes the \gls{O1} search results with a plot of the background distribution of the detection statistic in the case of \gls{O1} real noise with the detection statistic of the three events in \gls{O1} indicated by vertical dashed lines, which shows that the detection statistic for all the events in \gls{O1} are within the background. Therefore, we conclude that no significant evidence was found to support the detection of gravitational-wave echoes in \gls{O1}.

\begin{table}[h!]
\begin{center}
\begin{ruledtabular}
\begin{tabular}{cd{3.5}d{3.5}c}
\multicolumn{1}{p{0.25\columnwidth}}{\centering{Event}} & \multicolumn{1}{p{0.25\columnwidth}}{\centering{Detection statistic}} & \multicolumn{1}{p{0.25\columnwidth}}{\centering{$p$-value}} & \multicolumn{1}{p{0.25\columnwidth}}{\centering{Statistical significance ($\sigma$)}}  \\ \hline
GW150914 & -1.3 & 0.806 & $< 1$  \\
GW151012 & 0.4 & 0.0873 & $1.4$  \\
GW151226 & -0.2 & 0.254 & $< 1$  \\     
\end{tabular}
\end{ruledtabular}
\caption{\label{Table: detection statistics for O1} The detection statistic and its corresponding statistical significance and $p$-value for the three events in Advanced LIGO's \gls{O1} data. None of the events have a detection statistic greater than the threshold for $5\sigma$ detection. The ordering of the events in their statistical significance is consistent with that reported by Nielsen \etal\cite{2018arXiv181104904N}.}
\end{center}
\end{table}

\begin{figure}[h]
\includegraphics[width=1.0\columnwidth]{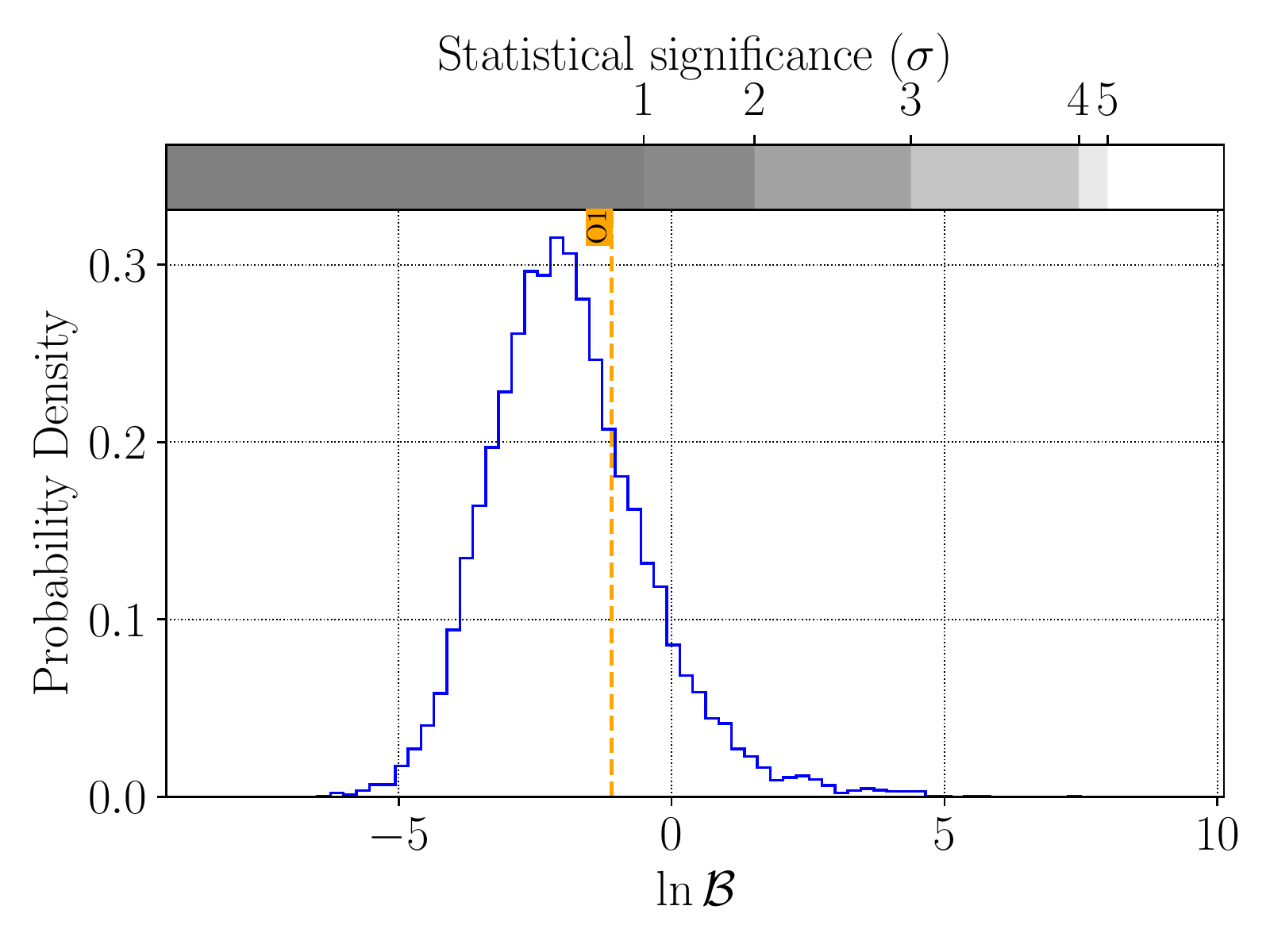}
\caption{\label{Fig: O1 results combined}The histogram of the sampled null distribution for the \emph{catalog log Bayes factor} $\ln \mathcal{B}^{\text{(cat)}}$ for Advanced LIGO's \gls{O1} data. This plot is similar to Fig. \ref{Fig: Catalog} but with a catalog of size $N_{\text{cat}} = 3$. From the figure, we see that the value of the catalog log Bayes factor for O1 is also well within the background, with a statistical significance $< 1\sigma$. Hence, we conclude that we also find no statistically significant combined evidence for the existence of gravitational-wave echoes in \gls{O1} data.}
\end{figure}

Apart from making statistical statements on individual events regarding whether gravitational-wave echoes are present or not, we can also make a statistical statement about whether gravitational-wave echoes are present in a collection of events (a catalog) as described in Sec. \ref{Subsec: Combine evidence}. The catalog log Bayes factor, which is the sum of the log Bayes factors for the three events in \gls{O1}, was found to be
\[
\ln \mathcal{B}^{\text{(cat)}}_{\text{O1}} = -1.1.
\]
Figure \ref{Fig: O1 results combined} shows the sampled background distribution for the catalog log Bayes factor for \gls{O1} noise, and the vertical dashed line indicates the catalog log Bayes factor found for the case of \gls{O1}. We see that the detected value for \gls{O1} does not stand out from the background, with a statistical significance $< 1\sigma$. This is expected as the log Bayes factors for individual events lie well within in the background. Hence, we conclude that we also find no statistically significant combined evidence for the existence of gravitational-wave echoes in \gls{O1} data.

\section{Discussions}
\subsection{Background and foreground distribution estimation}\label{Subsec: Discussion on injection sets}
During background and foreground estimation, the values of the \gls{IMR} parameters were not fixed to the same as those of a particular gravitational-wave event but rather were drawn from distributions that are representative of what Advanced \gls{LIGO} and Advanced Virgo would detect. It is legitimate to do this because in our hypotheses (see Sec. \ref{Subsec: Bayesian hypothesis testing}) we did not require the \gls{IMR} parameters to be known a priori. By allowing the \gls{IMR} parameters to be different during background and foreground estimation, our sampled foreground distribution can be used to estimate the efficiency of our search to detect gravitational-wave echoes for a variety of \acrlong{IMRE} signals. Similarly, our estimated background distribution can be used to estimate the false alarm probability for a variety of \acrlong{IMR} signals. Although the background and foreground estimation can also be done specifically for each individual gravitational-wave event, it will soon become computationally too expensive as we will have more than ten binary black hole mergers in an observing run; for example in the third observing run of Advanced \gls{LIGO} and Virgo, it was estimated that there will be about $35^{+78}_{-26}$ binary black hole merger events \cite{O3RateEstimate}. One can perform a follow-up analysis on interesting echo triggers found in this search to obtain a more accurate estimate of the false alarm probability and hence statistical significance.

\subsection{Combining evidence from multiple gravitational-wave echoes events}
As described in Sec. \ref{Subsec: Combine evidence}, we can combine Bayesian evidence by simply multiplying the Bayes factors from each independent event to give the catalog Bayes factor. Note that when combining evidence, we do not assume GW events are described by the same set of echo parameters, whereas in Abedi \etal{} they assumed that each event has the same values of $\gamma$ and $t_{0}/\overline{\Delta t_{\text{echo}}}$, where $\overline{\Delta t_{\text{echo}}}$ denotes the average value of $\Delta t_{\text{echo}}$ inferred in the events, which may not be the case in reality \cite{2017PhRvD..96h2004A}. In addition, they combined the GW events by summing up the $\rho^2$'s of each event, without demonstrating that this is a proper way to combine multiple measurements.

Combining Bayesian evidence from multiple gravitational-wave echoes events can provide tighter constraints on the existence of gravitational-wave echoes than a single event. Note that the null distribution for the catalog Bayes factor (and hence catalog log Bayes factor) can be constructed from the null distributions of log Bayes factors in a catalog of events. 

\subsection{Comparisons with other proposed search methodologies for gravitational-wave echoes}
\subsubsection{Model-dependent method proposed by Abedi et al.~}
As mentioned in the Introduction, Abedi \etal{} proposed a template-based search methodology for echoes using matched filtering. The parameter estimation was achieved by maximizing the square of the signal-to-noise ratio $\rho^2$, which is also defined in Eq. \ref{Eq: SNR}. The set of echo parameters that give the highest value of $\rho^2$ were said to be the inferred values in their analysis. Also, they used $\rho^2$ as the detection statistic of their search. By finding the number of events, in segments of data without gravitational waves, that have a higher or equal value of the detection statistic found in a candidate, the background distribution of their detection statistic can be estimated.

However, the use of $\rho^2$ as the detection statistic is suboptimal because a large short-time instrumental noise fluctuation (also known as a glitch) can easily cause a peak in $\rho^2$, and as a result the search will be trying to overfit the glitch instead of echoes. Also, the addition of five echo parameters when searching for echoes in gravitational-wave data will often make IMRE templates fit the data better than IMR templates when using $\rho^2$ as the detection statistic as there are more free parameters to be adjusted to fit the noise in the data. The Occam factor (described in Sec. \ref{Subsubsec: Occam's factor}) embodied in Bayesian model selection can mitigate the problem described above by penalizing more complicated models (i.e. having more free parameters), making our choice of the log Bayes factor more robust against noise than $\rho^2$ as chosen by Abedi \etal{}\cite{2017PhRvD..96h2004A}.

\subsubsection{Model-dependent Bayesian model selection approach proposed by Nielsen et al.} \label{Subsubsec: AEI Inference}
Nielsen \etal{} adopted the same Bayesian model selection framework to search for the existence of gravitational-wave echoes. In their work, they considered two hypotheses -- gravitational-wave strain data consisting of both echoes plus Gaussian noise and data consisting only of Gaussian noise -- and they are only looking at the post-merger part of a confirmed gravitational-wave signal. In comparison to our work, we select between two hypotheses: gravitational-wave strain data consisting of an \acrlong{IMRE} signal plus Gaussian noise versus an \acrlong{IMR} signal plus Gaussian noise model. Although both works are concerned with whether the data contain echoes, the hypotheses that are tested are not equivalent.

They interpreted their detection statistic (log Bayes factor) in the Bayesian way that when it is greater than zero, the data favor the echo signal plus Gaussian noise hypothesis and when it is less than zero, the data favor the pure Gaussian noise hypothesis. However, it is known and also pointed out by Nielsen \etal{} that the noise in real gravitational-wave strain data is not strictly Gaussian, and thus the aforementioned interpretation is only approximately true. We can also see from Fig. \ref{Fig: Null distribution} that the background distribution for our detection statistic, the log Bayes factor of \gls{IMRE} versus \gls{IMR}, is different for the cases of Gaussian noise and real noise in \gls{O1}, with a noticeable tail to the right in the case of \gls{O1}.

When interpreting the O1 results, they stated that a log Bayes factor with a value less than 1 does ``not worth more than a bare mention'' \cite{2018arXiv181104904N}. The use of a nomenclature to interpret the (log) Bayes factor is suboptimal because the scale to interpret the Bayesian evidence is not universally applicable. As mentioned by Kass \etal{} \cite{doi:10.1080/01621459.1995.10476572}, for forensic evidence to be ``conclusive'' in court trials, the Bayes factor needs to be at least 10 times larger than what was originally suggested by Jeffreys, which suggests that the scale is not universally applicable to different situations. A rigorous justification that the scale proposed by Kass \etal{}\cite{doi:10.1080/01621459.1995.10476572} is appropriate was not given in that paper, and its authors merely mentioned that ``From our own experience, these categories seem to furnish appropriate guidelines.'' When performing the parameter estimation with a Gaussian noise model, the effects due to the non-Gaussianity of the noise can only be properly accounted by sampling the background distribution of the log Bayes factor in real data, which was done in our work as described in Sec. \ref{Subsec: Background estimation}.

When generating templates of gravitational-wave echoes for parameter estimation, Nielsen et al.~chose to fix the parameters that govern the \acrlong{IMR} part of the waveform such as the component masses and the luminosity distance of the source of the signal. However, these inferred parameters have non-negligible uncertainties. By allowing the \gls{IMR} parameters to vary during parameter estimation as we do in our work, we can marginalize over these parameters in model selection properly and hence get a more accurate value for the log Bayes factor, instead of replacing the joint posterior distribution, that carries information such as the correlation between parameters, with a product of Dirac delta functions.

\subsubsection{Model-agnostic method proposed by Tsang et al.~}
Tsang \etal{}~\cite{2018PhRvD..98b4023T} also adopted the Bayesian approach and used the log Bayes factor as the detection statistic. However, their search methodology is morphology agnostic, meaning that no detailed knowledge of the waveform of gravitational-wave echoes is needed prior to a search. Their method was a modified version of the search pipeline \texttt{BayesWave} used for searching gravitational-wave bursts, which is suitable for searching gravitational-wave signals that are unmodeled or poorly modeled. This is exactly the current status of the modeling of gravitational-wave echoes emitted from exotic compact objects, where there is no consensus that a particular waveform model can accurately model echoes. However, it is exactly because their search methodology requires no prior knowledge on the waveform of gravitational-wave echoes that echoes need to be loud in order to be detected by their search. As in our proposed search methodology, we can make use of the knowledge on the waveform to extract weaker echoes buried in noise and the estimation of physical parameters of echoes can be done easily.

\section{Conclusions and Future Work}
In this \this{}, we have demonstrated that our proposed search methodology using Bayesian model selection between the presence of echoes and their absence can identify and estimate the parameters of an \gls{IMRE} signal buried in both Gaussian noise and real noise in the \gls{O1} data of Advanced \gls{LIGO}. In the validation test, the recovered echo parameters were both close to the true value and had narrow posterior probability distributions. We demonstrated that we can use a Bayesian model selection to test the existence of echoes in simulated data, and report the statistical significance of the detection. By performing many analyses on simulated data with gravitational-wave echoes injected, we also found that the search was able to identify gravitational-wave echoes in simulated Gaussian noise about \GaussianEfficiencyPercentage{}\% of the time and in \gls{O1} real noise about \OOneEfficiencyPercentage{}\% of the time with $\geq 5\sigma$ significance. Applying the search methodology to search for gravitational-wave echoes in the three \gls{O1} events, we found no statistically significant evidence to support the detection of echoes.

In the future, we can repeat the analysis with different parametrized gravitational-wave echoes waveform models that are more physical to provide more realistic evidence of the existence of echoes from \acrlong{ECO}s. When we understand the physics of \acrlong{ECO}s better in the sense that we can come up with physical waveform models of the echoes from different types of \acrlong{ECO}s, the methodology proposed in this paper can be readily modified to test the nature of \acrlong{ECO}s, using subhypotheses of $\mathcal{H}_{1}$ such as $\mathcal{H}_{\text{Gravastar}}$ and $\mathcal{H}_{\text{Fuzzball}}$, that is
\[ \mathcal{H}_{1} = \mathcal{H}_{\text{Gravastar}} \vee \mathcal{H}_{\text{Fuzzball}} \vee \ldots,\]
so that we can learn even more about the properties and structure of \acrlong{ECO}s.

\begin{acknowledgements}
The authors acknowledge the generous support from the National Science Foundation in the United States. \gls{LIGO} was constructed by the California Institute of Technology and Massachusetts Institute of Technology with funding from the National Science Foundation and operates under cooperative agreement PHY-0757058. Virgo is funded by the French Centre National de Recherche Scientifique (CNRS), the Italian Istituto Nazionale della Fisica Nucleare (INFN) and the Dutch Nikhef, with contributions by Polish and Hungarian institutes. R.K.L.L. and T.G.F.L. would also like to gratefully acknowledge the support from the Croucher Foundation in Hong Kong. The work described in this \this{} was partially supported by a grant from the Research Grants Council of the Hong Kong (Project No. CUHK 24304317) and the Direct Grant for Research from the Research Committee of the Chinese University of Hong Kong.
The authors acknowledge the use of the IUCAA LDG cluster Sarathi for the computational/numerical work. The authors are also grateful for computational resources provided by the LIGO Laboratory and supported by National Science Foundation Grants PHY-0757058 and PHY-0823459. 
The authors would like to thank Ajith Parameswaran for reviewing this \this{} during the LSC internal review.
R.K.L.L. would also like to thank Zachary Mark, Rory Smith, Peter T.~H.~Pang, Ignacio Magana, Alex Nielsen, Ofek Birnholtz and Yanbei Chen for the fruitful conversations with the first author. Figure \ref{Fig: Corner} was generated using the Python package \verb|corner.py| \cite{corner}.
This paper carries LIGO Document Number LIGO-\DocumentID{}.
\end{acknowledgements}

\appendix

\clearpage
\bibliographystyle{apsrev4-1}
\bibliography{database}%

\end{document}